\documentclass[12pt]{article}
\usepackage[utf8]{inputenc}
\usepackage[T1]{fontenc}
\usepackage{theorem}
\usepackage{bbm}
\usepackage{mathtools}
\usepackage{multirow}
\usepackage{amssymb}
\usepackage{array}
\usepackage{color}
\usepackage{graphicx}
\usepackage{pgfplots}
\usepackage{tikz}
\usetikzlibrary{positioning}
\usepackage[export]{adjustbox}
\usetikzlibrary{shapes.geometric, arrows}
\usepackage{amsmath}
\usepackage{verbatim} 
\usepackage{rotating}
\usepackage[round]{natbib}
\usepackage[colorlinks=true,linkcolor=black, citecolor=blue, urlcolor=blue]{hyperref}
\usepackage{footnote}
\usepackage{tcolorbox}
\makesavenoteenv{tabular}
\makesavenoteenv{table}

\newtheorem{theorem}{Theorem}
\newtheorem{lemma}{Lemma}

\newtheorem{algorithm}{Algorithm}
\newtheorem{proofprop}{\textit{Proof of Proposition}}
\newtheorem{remark}{Remark}
\newtheorem{prop}{Proposition}
\newtheorem{prooflemma}{\textit{Proof of Lemma}}
\newtheorem{proofremark}{\textit{Proof of Remark}}
\newcommand\eqdef{\stackrel{\mathclap{\normalfont\mbox{def}}}{=}}
\newcommand*{\QED}{\hfill\ensuremath{\blacksquare}}
\DeclareMathOperator*{\argmin}{argmin} 
\newtheorem{propA}{Proposition}

\newtheorem{proofA}{\textit{Proof of Proposition}}

\newtheorem{proofthm}{Proof of Theorem}

\newtheorem{model}{\textbf{Model}}

\newcommand\tab[1][1cm]{\hspace*{#1}}
\begin{document}

\begin{center}
	\Large{{\scshape A Kernel-based Consensual Aggregation for Regression}}\\
	
	\bigskip
	
	\normalsize
	Sothea Has
\end{center}

\begin{flushleft}
LPSM, Sorbonne Université Pierre et Marie Curie (Paris 6)\\ 75005 Paris, France\\
\url{sothea.has@lpsm.paris}
\end{flushleft}

\begin{abstract}
%In this article, we introduce a kernel-based consensual aggregation method for regression problems. We aim to flexibly combine individual regression estimators $r_1,r_2,...,r_M$ using a weighted average where the weights are defined based on some kernel function in order to build a target prediction. It may be seen as a kernel smoother method implemented on the features of predictions, given by all the individual estimators, instead of the original inputs. This work extends the context of \cite{cobra} to a more general kernel-based framework. We show that this configuration asymptotically inherits the consistency property of the basic consistent estimators. Moreover, we propose to numerically learn the key parameter of the method using a gradient descent algorithm instead of using the grid search one. The numerical experiments carried out on several simulated and real datasets suggest that the performance of the method is improved with the introduction of smoother kernel functions.
In this article, we introduce a kernel-based consensual aggregation method for regression problems. We aim to flexibly combine individual regression estimators $r_1, r_2, ..., r_M$ using a weighted average where the weights are defined based on some kernel function to build a target prediction. This work extends the context of \cite{cobra} to a more general kernel-based framework. We show that this more general configuration also inherits the consistency of the basic consistent estimators. Moreover, an optimization method based on gradient descent algorithm is proposed to efficiently and rapidly estimate the key parameter of the strategy. The numerical experiments carried out on several simulated and real datasets are also provided to illustrate the speed-up of gradient descent algorithm in estimating the key parameter and the improvement of overall performance of the method with the introduction of smoother kernel functions.
\end{abstract}

\noindent \emph{Keywords:}
Consensual aggregation, kernel, regression.

\bigskip

\noindent \emph{2010 Mathematics Subject Classification:} {62G08, 62J99, 62P30}

\section{Introduction}
Aggregation methods, given the high diversity of available estimation strategies, are now of great interest in constructing predictive models. To this goal, several aggregation methods consisting of building a linear or convex combination of a bunch of initial estimators have been introduced, for instance in \cite{Cat04}, \cite{JN00}, \cite{Nem00}, \cite{yang2000,yang2001,yang2004}, \cite{bookDistributionFree}, \cite{W03}, \cite{Aud}, \cite{BTW06,BTW07a,BTW07b}, and \cite{DalTsy}. Another approach of model selection, which aims at selecting the best estimator among the candidate estimators, has also been proposed (see, for example, \cite{MassStF}).

Apart from the usual linear combination and model selection methods, a different technique has been introduced in classification problems by \cite{mojirsheibani1999}. In his paper, the combination is the \textit{majority vote} among all the points for which their predicted classes, given by all the basic classifiers, \textit{coincide} with the predicted classes of the query point. Roughly speaking, instead of predicting a new point based on the structure of the original input, we look at the topology defined by the predictions of the candidate estimators. Each estimator was constructed differently so may be able to capture different features of the input data and useful in defining ``closeness''. Consequently, two points having similar predictions or classes seem reasonably having similar actual response values or belonging to the same actual class.

Later, \cite{mojirsheibani2000} and \cite{majidAndKong2016} introduced exponential and general kernel-based versions of the primal idea to improve the smoothness in selecting and weighting individual data points in the combination. In this context, the kernel function transforms the level of \textit{disagreements} between the predicted classes of a training point $x_i$ and the query point $x$ into a contributed weight given to the corresponding point in the vote. Besides, \cite{cobra} configured the original idea of \cite{mojirsheibani1999} as a regression framework where a training point $x_i$ is ``close'' to the query point $x$ if each of their predictions given by all the basic regression estimators is ``close''. Each of the close neighbors of $x$ will be given a uniformly 0-1 weight contributing to the combination. It was shown theoretically in these former papers that the combinations inherit the consistency property of consistent basic estimators.

Recently from a practical point of view, a kernel-based version of \cite{cobra} called \texttt{KernelCobra} has been implemented in \texttt{pycobra} python library (see \cite{pycobra}). Moreover, it has also been applied in filtering to improve the image denoising (see \cite{imageDenoisng}). In a complementary manner to the earlier works, we present another kernel-based consensual regression aggregation method in this paper, as well as its theoretical and numerical performances. More precisely, we show that the consistency inheritance property shown in \cite{cobra} also holds for this kernel-based configuration for a broad class of regular kernels. Moreover, an evidence of numerical simulation carried out on a similar set of simulated models, and some real datasets shows that the present method outperforms the classical one.

This paper is organized as follows. Section~\ref{sec:theorem} introduces some notation, the definition of the proposed method, and presents the theoretical results, namely consistency and convergence rate of the variance term of the method for a subclass of regular kernel functions. A method based on gradient descent algorithm to estimate the bandwidth parameter is described in Section~\ref{sec:optimization}. Section~\ref{sec:numeric} illustrates the performances of the proposed method through several numerical examples of simulated and real datasets. Lastly, Section~\ref{sec:proof} collects all the proofs of the theoretical results given in Section~\ref{sec:theorem}.

\section{The kernel-based combining regression}
\label{sec:theorem}
\subsection{Notation}
We consider a training sample $\mathcal{D}_n=\{(X_i,Y_i)_{i=1}^n\}$ where $(X_i,Y_i), i=1,2,...,n$, are {\it iid} copies of the generic couple $(X,Y)$. We assume that $(X,Y)$ is an $\mathbb{R}^d\times\mathbb{R}$-valued random variable with a suitable integrability which will be specified later. 

We randomly split the training data $\mathcal{D}_n$ into two parts of size $\ell$ and $k$ such that $\ell+k=n$, which are denoted by $\mathcal{D}_{\ell}=\{(X_i^{(\ell)},Y_i^{(\ell)})_{i=1}^{\ell}\}$ and $\mathcal{D}_k=\{(X_i^{(k)},Y_i^{(k)})_{i=1}^{k}\}$ respectively (a common choice is $k=\lceil n/2\rceil=n-\ell$). The $M$ basic regression estimators or machines $r_{k,1},r_{k,2},...,r_{k,M}$ are constructed using only the data points in $\mathcal{D}_k$. These basic machines can be any regression estimators such as linear regression, $k$NN, kernel smoother, SVR, lasso, ridge, neural networks, naive Bayes, bagging, gradient boosting, random forests, etc. They could be parametric, nonparametric or semi-parametric with their possible tuning parameters. For the combination, we only need the predictions given by all these basic machines of the remaining part $\mathcal{D}_{\ell}$ and  the query point $x$. 

In the sequel, for any $x\in\mathbb{R}^d$, the following notation is used:
\begin{itemize}
\item$\textbf{r}_k(x)=(r_{k,1}(x),r_{k,2}(x),...,r_{k,M}(x))$: the vector of predictions of $x$.
\item$\|x\|=\|x\|_2=\sqrt{\sum_{i=1}^dx_i^2}$: Euclidean norm on $\mathbb{R}^d$.
\item$\|x\|_1=\sum_{i=1}^d|x_i|$: $\ell_1$ norm on $\mathbb{R}^d$.
\item$g^*(x)=\mathbb{E}[Y|X=x]$: the regression function.
\item$g^*(\textbf{r}_k(x))=\mathbb{E}[Y|\textbf{r}_k(x)]$: the conditional expectation of the response variable given all the predictions. This can be proven to be the optimal estimator in regression over the set of predictions $\textbf{r}_k(X)$.
\end{itemize}
The consensual regression aggregation is the weighted average defined by
\begin{equation}
\label{eq:combine}
g_n(\textbf{r}_k(x))=\sum_{i=1}^{\ell}W_{n,i}(x)Y_i^{(\ell)}.
\end{equation}

Recall that given all the basic machines $r_{k,1},r_{k,2},...,r_{k,M}$, the aggregation method proposed by \cite{cobra} corresponds to the following naive weights:

\begin{equation}
\label{eq:cobra1}
W_{n,i}(x)=\frac{\prod_{m=1}^M\displaystyle\mathbbm{1}_{\{|r_{k,m}(X_i)-r_{k,m}(x)|<h\}}}{\sum_{j=1}^{\ell}\prod_{m=1}^M\mathbbm{1}_{\{|r_{k,m}(X_j)-r_{k,m}(x)|<h\}}},i=1,2,...,\ell.
\end{equation}
Moreover, the condition of ``closeness for all'' predictions, can be relaxed to ``some'' predictions, which corresponds to the following weights:
\begin{equation}
\label{eq:cobra2}
W_{n,i}(x)=\frac{\displaystyle\mathbbm{1}_{\{\sum_{m=1}^M\mathbbm{1}_{\{|r_{k,m}(X_i)-r_{k,m}(x)|<h\}}\geq \alpha M\}}}{\sum_{j=1}^{\ell}\mathbbm{1}_{\{\sum_{m=1}^M\mathbbm{1}_{\{|r_{k,m}(X_j)-r_{k,m}(x)|<h\}}\geq \alpha M\}}},i=1,2,...,\ell
\end{equation}
where $\alpha\in\{1/M,2/M,...,1\}$ is the proportion of consensual predictions required and $h>0$ is the bandwidth or window parameter to be determined. Constructing the proposed method is equivalent to searching for the best possible value of these parameters over a given grid, minimizing some quadratic error which will be described in Section~\ref{sec:optimization}.

In the present paper, $K:\mathbb{R}^M\to\mathbb{R}_+$ denotes a regular kernel which is a decreasing function satisfying:
\begin{align}
\label{eq:regular}
\exists b,\kappa_0,\rho>0\ \text{such that}
\begin{cases}
b\mathbbm{1}_{B_M(0,\rho)}(z)\leq K(z)\leq 1, \forall z\in\mathbb{R}^M\\
			\int_{\mathbb{R}^M}\sup_{u\in B_M(z,\rho)}K(u)dz = \kappa_0 < +\infty
		\end{cases}
	\end{align}
	where $B_M(c,r)=\{z\in\mathbb{R}^M:\|c-z\|<r\}$ denotes the open ball of center $c\in\mathbb{R}^M$ and radius $r>0$ of $\mathbb{R}^M$. We propose in \eqref{eq:combine} a method associated to the weights defined at any query point $x\in\mathbb{R}^d$ by
\begin{align}
\label{eq:KCOBRA}
W_{n,i}(x)=\frac{K_h(\textbf{r}_k(X_i^{(\ell)})-\textbf{r}_k(x))}{\sum_{j=1}^{\ell}K_h(\textbf{r}_k(X_j^{(\ell)})-\textbf{r}_k(x))},i=1,2,...,\ell %+\frac{\mathbbm{1}_{\{\Sigma(x)=0\}}}{\ell}
\end{align}
where $K_h(z)=K(z/h)$ for some bandwidth parameter $h>0$ with the convention of $0/0=0$. Observe that the combination is based only on $\mathcal{D}_{\ell}$ but the whole construction of the method depends on the whole training data $\mathcal{D}_n$ as the basic machines are all constructed using $\mathcal{D}_k$. In our setting, we treat the vector of predictions $\textbf{r}_k(x)$ as an $M$-dimensional feature, and the kernel function is applied on the whole vector at once. Note that the implementation of \texttt{KernelCobra} in \cite{Guedj_2020} corresponds to the following weights:
\begin{equation}
\label{eq:Guedj}
W_{n,i}(x)=\frac{\sum_{m=1}^MK_h(r_{k,m}(X_i^{(\ell)})-r_{k,m}(x))}{\sum_{j=1}^{\ell}\sum_{m=1}^MK_h(r_{k,m}(X_j^{(\ell)})-r_{k,m}(x))},i=1,2,...,\ell
\end{equation}
where the univariate kernel function $K$ is applied on each component of $\textbf{r}_k(x)$ separately.
\subsection{Theoretical performance}
The performance of the combining estimation $g_n$ is measured using the quadratic risk defined by
\begin{align*}
\mathbb{E}\Big[|g_n(\textbf{r}_k(X))-g^*(X)|^2\Big]
\end{align*}
where the expectation is taken with respect to both $X$ and the training sample $\mathcal{D}_n$. Firstly, we begin with a simple decomposition of the distortion between the proposed method and the optimal regression estimator $g^*(X)$ by introducing the optimal regression estimator over the set of predictions $g^*(\textbf{r}_k(X))$. The following proposition shows that the nonasymptotic-type control of the distortion, presented in Proposition.2.1 of \cite{cobra}, also holds for this case of regular kernels.

\begin{prop}
\label{prop:1}
Let $\textbf{r}_k=(r_{k,1},r_{k,2},...,r_{k,M})$ be the collection of all basic estimators and $g_n(\textbf{r}_k(x))$ be the combined estimator defined in \eqref{eq:combine} with the weights given in \eqref{eq:KCOBRA} computed at point $x\in\mathbb{R}^d$. Then, for all distributions of $(X,Y)$ with $\mathbb{E}[|Y|^2]< +\infty$,
\begin{align*}
	\mathbb{E}\Big[|g_n(\textbf{r}_k(X))-g^*(X)|^2\Big]&\leq \inf_{f\in\mathcal{G}}\mathbb{E}\Big[|f(\textbf{r}_k(X))-g^*(X)|^2\Big]\\
	&\quad+\mathbb{E}\Big[|g_n(\textbf{r}_k(X))-g^*(\textbf{r}_k(X))|^2\Big]
\end{align*}
where $\mathcal{G}$ is the class of any function $f:\mathbb{R}^M\to\mathbb{R}$ satisfying $\mathbb{E}[f(\textbf{r}_k(X))|^2]<+\infty$. In particular,
	 \begin{align*}
	\mathbb{E}\Big[|g_n(\textbf{r}_k(X))-g^*(X)|^2\Big]&\leq \min_{1\leq m\leq M}\mathbb{E}\Big[|r_{k,m}(X)-g^*(X)|^2\Big]\\
	&\quad+\mathbb{E}\Big[|g_n(\textbf{r}_k(X))-g^*(\textbf{r}_k(X))|^2\Big].
\end{align*}
\end{prop}

%Proposition~\ref{prop:1} ensures the performance of $g_n$ with respect to the performance of best basic regression estimator and the variation of the combination itself for whatever the distribution of $(X,Y)$ such that $Y$ is square-integrable. 

The two terms of the last bound can be viewed as a bias-variance decomposition where the first term $\min_{1\leq m\leq M}\mathbb{E}[|r_{k,m}(X)-g^*(X)|^2]$ can be seen as the bias and $\mathbb{E}[|g_n(\textbf{r}_k(X))-g^*(\textbf{r}_k(X))|^2]$ is the variance-type term (\cite{cobra}). Given all the machines, the first term cannot be controlled as it depends on the performance of the best constructed machine, and it will be there as the asymptotic control of the performance of the proposed method. Our main task is to deal with the second term, which can be proven to be asymptotically negligible in the following key proposition.

\begin{prop}
\label{prop:2}
	Assume that $r_{k,m}$ is bounded for all $m=1,2,..,M$. Let $h\rightarrow0$ and $\ell\rightarrow+\infty$ such that $h^M\ell\to+\infty$. Then
	\begin{align*}
	\mathbb{E}\Big[|g_n(\textbf{r}_k(X))-g^*(\textbf{r}_k(X))|^2\Big]\rightarrow0\ \text{as }\ell\rightarrow+\infty
	\end{align*}
	for all distribution of $(X,Y)$ with $\mathbb{E}[|Y|^2]<+\infty$. Thus,
	\begin{align*}
	\limsup_{\ell\rightarrow+\infty}\mathbb{E}\Big[|g_n(\textbf{r}_k(X))-g^*(X)|^2\Big]\leq\inf_{f\in\mathcal{G}}\mathbb{E}\Big[|f(\textbf{r}_k(X))-g^*(X)|^2\Big].
	\end{align*}
	And in particular,
	\begin{align*}
	\limsup_{\ell\rightarrow+\infty}\mathbb{E}\Big[|g_n(\textbf{r}_k(X))-g^*(X)|^2\Big]\leq\min_{1\leq m \leq M}\mathbb{E}\Big[|r_{k,m}(X)-g^*(X)|^2\Big].
	\end{align*}
\end{prop}
Proposition~\ref{prop:2} above is an analogous setup of Proposition 2.2 in \cite{cobra}. To prove this result, we follow the procedure of Stone's theorem (see, for example, \cite{stone1977} and Chapter 4 of \cite{bookDistributionFree}) of weak universal consistency of non-parametric regression. However, showing this result for the class of regular kernels is not straightforward. Most of the previous studies provided such a result of $L_2$-consistency only for the class of compactly supported kernels (see, for example, Chapter 5 of \cite{bookDistributionFree}). In this study, we can derive the result for this broader class thanks to the boundedness of all basic machines. However, the price to pay for the universality for this class of regular kernels is the lack of convergence rate. To this goal, a weak smoothness assumption of $g^*$ with respect to the basic machines is required. For example, the convergence rate of the variance-type term in \cite{cobra} is of order $O(\ell^{-2/(M+2)})$ under the same smoothness assumption, and this result also holds for all the compactly support kernels. Our goal is not to theoretically do better than the classical method but to investigate such a similar result in a broader class of kernel functions. For those kernels which the tails decrease fast enough, the convergence rate of the variance-type term can be attained as described in the following main theorem of this paper.

\begin{theorem}
\label{thm.1}
Assume that the response variable $Y$ and all the basic machines $r_{k,m},m=1,2,...,M$, are bounded by some constant $R$. Suppose that there exists a constant $L\geq0$ such that, for every $k\geq1$,
\begin{equation*}
|g^*(\textbf{r}_k(x))-g^*(\textbf{r}_k(y))|\leq L\|\textbf{r}_k(x)-\textbf{r}_k(y)\|,\forall x,y\in\mathbb{R}^d.
\end{equation*}
We assume moreover that
\begin{equation}
	\label{eq:assumption}
	\exists R_K,C_k>0:K(z)\|z\|^2\leq\frac{C_K}{1+\|z\|^{M}}, \forall z\in\mathbb{R}^M\ \text{such that }\|z\|\geq R_K.
\end{equation}
Then, with the choice of $h\propto \ell^{-\frac{M+2}{M^2+2M+4}}$, one has
\begin{equation}
\label{eq:convergRate}
\mathbb{E}[|g_n(\textbf{r}_k(X))-g^*(X)|^2]\leq \min_{1\leq m\leq M}\mathbb{E}[|r_{k,m}(X)-g^*(X)|^2]+C\ell^{-\frac{4}{M^2+2M+4}}
\end{equation}
for some positive constant $C=C(b,L,R,R_K,C_K)$ independent of $\ell$. 
\end{theorem} 
Moreover, if there exists a consistent estimator named $r_{k,m_0}$ among $\{r_{k,m}\}_{m=1}^M$ i.e.,
$$\mathbb{E}[|r_{k,m_0}(X)-g^*(X)|^2]\to0\ \ \text{as }k\to+\infty,$$
then the combing estimator $g_n$ is also consistent for all distribution of ($X,Y$) in some class $\mathcal{M}$. Consequently, under the assumption of Theorem~\ref{thm.1}, one has 
$$\lim_{k,\ell\to+\infty}\mathbb{E}[|g_n(\textbf{r}_k(X))-g^*(X)|^2]=0.$$

\begin{remark}
\label{rmk:boundK}
The assumption on the upper bound of the kernel $K$ in the theorem above is very weak, chosen so that the result holds for a large subclass of regular kernels. However, the convergence rate is indeed slow for this subclass of kernel functions. If we strengthen this condition, we can obtain a much nicer result. For instance, if we assume that the tails decrease at least of exponential speed i.e.,
$$\exists R_K,C_K>0\text{ and }\alpha\in(0,1):K(z)\leq C_Ke^{-\|z\|^{\alpha}},\forall z\in\mathbb{R}^M,\|z\|\geq R_K,$$ by following the same procedure as in the proof of the above theorem (Section~\ref{sec:proof}), one can easily check that the convergence rate of the variance-type term is of order $O(\ell^{-2\alpha/(M+2\alpha)})$. This rate approaches the state of the art of the classical method by \cite{cobra} when $\alpha$ approaches $1$.
\end{remark}

\section{Bandwidth parameter estimation thanks to gradient descent}
\label{sec:optimization}

In earlier works by \cite{cobra} and \cite{Guedj_2020}, the training data $\mathcal{D}_{n}$ is practically broken down into three parts $\mathcal{D}_k$ where all the candidate machines $\{\textbf{r}_{k,m}\}_{m=1}^M$ are built, and two other parts $\mathcal{D}_{\ell_1}$ and $\mathcal{D}_{\ell_2}$. $\mathcal{D}_{\ell_1}$ is used for the combination defined in equation~(\ref{eq:combine}), and $\mathcal{D}_{\ell_2}$ is the validation set used to learn the bandwidth parameter $h$ of equation~(\ref{eq:cobra1}) and the proportion $\alpha$ of equation~(\ref{eq:cobra2}) by minimizing the average quadratic error evaluated on $\mathcal{D}_{\ell_2}$ defined as follows,
\begin{align}
\label{eq:error1}
\varphi_M(h)=\frac{1}{|\mathcal{D}_{\ell_2}|}\sum_{(X_j,Y_j)\in\mathcal{D}_{\ell_2}}[g_n(\textbf{r}_k(X_j))-Y_j]^2
\end{align}
where $|\mathcal{D}_{\ell_2}|$ denotes the cardinality of $\mathcal{D}_{\ell_2}$, $g_n(\textbf{r}_k(X_j))=\sum_{(X_i,Y_i)\in\mathcal{D}_{\ell_1}}W_{n,i}(X_j)Y_i$ defined in equation~(\ref{eq:combine}), and the weight $W_{n,i}(X_j)$ is given in equation~\eqref{eq:cobra1} and \eqref{eq:Guedj} for \cite{cobra} and \cite{Guedj_2020} respectively. Note that the subscript $M$ of $\varphi_M(h)$ indicates the full consensus between the $M$ components of the predictions $\textbf{r}_k(X_i)$ and $\textbf{r}_k(X_j)$ for any $X_i$ and $X_j$ of $\mathcal{D}_{\ell_1}$ and $\mathcal{D}_{\ell_2}$ respectively. In this case, constructing a combining estimation $g_n$ is equivalent to searching for an optimal parameter $h^*$ over a given grid $\mathcal{H}=\{h_{\min},...,h_{\max}\}$ i.e.,
$$h^*=\argmin_{h\in\mathcal{H}}\varphi_M(h).$$ 
The parameter $\alpha$ of equation~(\ref{eq:cobra2}) can be tuned easily by considering $\varphi_{\alpha M}(h)$ where $\alpha\in\{1/2,1/3,...,1\}$ referring to the proportion of consensuses required among the $M$ components of the predictions. In this case, the optimal parameters $\alpha^*$ and $h^*$ are chosen to be the minimizer of $\varphi_{\alpha M}(h)$ i.e.,
$$(\alpha^*,h^*)=\argmin_{(\alpha,h)\in\{1/2,1/3,...,1\}\times\mathcal{H}}\varphi_{\alpha M}(h).$$ 
Note that in both papers, the grid search algorithm is used in searching for the optimal bandwidth parameter. 

In this paper, the training data is broken down into only two parts, $\mathcal{D}_{k}$ and $\mathcal{D}_{\ell}$. Again, we construct the basic machines using $\mathcal{D}_{k}$, and we propose the following $\kappa$-fold cross-validation error which is a function of the bandwidth parameter $h>0$ defined by
\begin{align}
\label{eq:error2}
\varphi^{\kappa}(h)=\frac{1}{\kappa}\sum_{p=1}^{\kappa}\sum_{(X_j,Y_j)\in F_p}[g_n(\textbf{r}_k(X_j))-Y_j]^2
\end{align}
where in this case, $g_n(\textbf{r}_k(X_j))=\sum_{(X_i,Y_i)\in \mathcal{D}_{\ell}\setminus F_p}W_{n,i}(X_j)Y_i$, is computed using the remaining $\kappa-1$ folds of $\mathcal{D}_{\ell}$ leaving $F_p\subset\mathcal{D}_{\ell}$ as the corresponding validation fold. We often observe the convex-like curves of the cross-validation quadratic error on many simulations; and from this observation, we propose to use a gradient descent algorithm to estimate the optimal bandwidth parameter. The associated gradient descent algorithm used to estimate the optimal parameter $h^*$ is implemented as follows:
\begin{tcolorbox}
\begin{algorithm}{: Gradient descent for estimating $h^*$:}
\begin{enumerate}
\item Initialization: $h_0$, a learning rate $\lambda>0$, threshold $\delta>0$ and the maximum number of iteration $N$.
\item For $k=1,2,...,N$, \textbf{while} $\Big|\frac{d}{d h}\varphi^{\kappa}(h_{k-1})\Big|>\delta$ do: 
$$h_k\gets h_{k-1}-\lambda\frac{d}{d h}\varphi^{\kappa}(h_{k-1})$$
\item return $h_k$ violating the \textbf{while} condition or $h_N$ to be the estimation of $h^*$.
\end{enumerate}
\end{algorithm}
\end{tcolorbox}
\noindent From equation~\eqref{eq:error2}, for any $(X_j,Y_j)\in F_p$, one has
\small{
\begin{align*}
\frac{d}{d h}\varphi^{\kappa}(h)&=\frac{1}{\kappa}\sum_{p=1}^{\kappa}\sum_{(X_j,Y_j)\in F_p}2\frac{\partial}{\partial h}g_n(\textbf{r}_k(X_j))(g_n(\textbf{r}_k(X_j))-Y_j)
\end{align*}
}
where 
\small{
\begin{align*}
g_n(\textbf{r}_k(X_j))&=\frac{\sum_{(X_i,Y_i)\mathcal{D}_{\ell}\in\setminus F_p}Y_iK_h(\textbf{r}_k(X_j)-\textbf{r}_k(X_i))}{\sum_{(X_q,Y_q)\in\mathcal{D}_{\ell}\setminus F_p}K_h(\textbf{r}_k(X_j)-\textbf{r}_k(X_q))}\\
\Rightarrow\frac{\partial}{\partial h}g_n(\textbf{r}_k(X_j))&=\sum_{(X_i,Y_i),(X_q,Y_q)\in\mathcal{D}_{\ell}\setminus F_p}(Y_i-Y_q)\times\\
&\quad\ \frac{\frac{\partial}{\partial h}K_h(\textbf{r}_k(X_j)-\textbf{r}_k(X_i))K_h(\textbf{r}_k(X_j)-\textbf{r}_k(X_q))}{\Big[\sum_{(X_i,Y_i)\mathcal{D}_{\ell}\in\setminus F_p}K_h(\textbf{r}_k(X_j)-\textbf{r}_k(X_i))\Big]^2}.
\end{align*}
}

\noindent The differentiability of $g_n$ depends entirely on the kernel function $K$. Therefore, for suitable kernels, the implementation of the algorithm is straightforward. For example, in the case of Gaussian kernel $K_h(x)=\exp(-h\|x\|^2/(2\sigma^2))$ for some $\sigma>0$, one has
\small{
\begin{align*}
\frac{\partial}{\partial h}g_n(\textbf{r}_k(X_j))&=\sum_{(X_i,Y_i),(X_q,Y_q)\in\mathcal{D}_{\ell}\setminus F_p}(Y_q-Y_i)\|\textbf{r}_k(X_j)-\textbf{r}_k(X_i)\|^2\times\\
&\quad\frac{\exp\Big(-h(\|\textbf{r}_k(X_j)-\textbf{r}_k(X_i)\|^2+\|\textbf{r}_k(X_j)-\textbf{r}_k(X_q)\|^2)/(2\sigma^2)\Big)}{2\sigma^2\Big(\sum_{(X_q,Y_q)\notin F_p}\exp(-h\|\textbf{r}_k(X_j)-\textbf{r}_k(X_q)\|^2/(2\sigma^2))\Big)^2}.
\end{align*}
}

In our numerical experiment, the numerical gradient of (\ref{eq:error2}) can be computed efficiently and rapidly thanks to \texttt{grad} function contained in \texttt{pracma} library of \texttt{R} program (see \cite{grad}). We observe that the algorithm works much faster, and more importantly it does not require the information of the interval containing the optimal parameter as the grid search does. Most of the time, the parameter $h$ vanishing the numerical gradient of the objective function can be attained, leading to a good construction of the corresponding combining estimation method, as reported in the next section.

\section{Numerical examples}
\label{sec:numeric}
This section is devoted to numerical experiments to illustrate the performance of our proposed method. It is shown in \cite{cobra} that the classical method mostly outperforms the basic machines of the combination. In this experiment, we compare the performances of the proposed methods with the classical one and all the basic machines. Several options of kernel functions are considered. Most kernels are compactly supported on $[-1,1]$, taking nonzero values only on $[-1,1]$, except for the case of compactly supported Gaussian which is supported on $[-\rho_1,\rho_1]$, for some $\rho_1>0$. Moreover to implement the gradient descent algorithm in estimating the bandwidth parameter, we also present the results of non-compactly supported cases such as classical Gaussian and 4-exponential kernels. All kernels considered in this paper are listed in Table~\ref{tab:kernels}, and some of them are displayed (univariate case) in Figure~\ref{fig:1} below.

\begin{table}[h!]
\centering 
\small
\begin{tabular}{| l | l |}  % creating 10 columns
\hline                      % inserting double-line 
\tab[0.45cm] \textbf{Kernel} & \tab[1.3cm] \textbf{Formula}\\
\hline
\hline
Naive\footnote{The naive kernel corresponds to the method by \cite{cobra}.} & $K(x)=\prod_{i=1}^d\mathbbm{1}_{\{|x_i|\leq 1\}}$\\ 
\hline
Epanechnikov & $K(x)=(1-\|x\|^2)\mathbbm{1}_{\{\|x\|\leq 1\}}$\\
\hline 
Bi-weight & $K(x)=(1-\|x\|^2)^2\mathbbm{1}_{\{\|x\|\leq 1\}}$\\
\hline
Tri-weight & $K(x)=(1-\|x\|^2)^3\mathbbm{1}_{\{\|x\|\leq 1\}} $\\
\hline
%Triangular & $K(x)=(1-\|x\|_1)\mathbbm{1}_{\{\|x\|_1\leq 1\}}$\\
%\hline  
Compact-support Gaussian & $K(x)=\exp\{-\|x\|^2/(2\sigma^2)\}\mathbbm{1}_{\{\|x\|\leq \rho_1\}}, \sigma,\rho_1>0$\\
\hline                          % inserts single-line
Gaussian & $K(x)=\exp\{-\|x\|^2/(2\sigma^2)\}, \sigma>0$\\
\hline  
$4$-exponential & $K(x)=\exp\{-\|x\|^4/(2\sigma^4)\}, \sigma>0$\\
\hline
%Cauchy & $K(x)=1/(1+\|x\|^2)$\\
\hline  
\end{tabular}
\caption{Kernel functions used.}%
\label{tab:kernels}
\end{table}

\begin{figure}[h!]
\centering
\begin{tikzpicture}
\begin{axis}[
    axis lines = left,
    xlabel = $x$,
    ylabel = {$K(x)$},
    legend style={at={(1,1)},anchor=north west}, 
    legend cell align={left},
    width=10cm,height=6.5cm
]
\addplot [
    domain=-3.5:-1, 
    samples=100, 
    color=black,
    line width=1.5pt,
]
{0};
\addplot [
    domain=1:3.5, 
    samples=100, 
    color=black,
    line width=1.5pt,
]
{0};

\addplot [
    domain=-1:1, 
    samples=100, 
    color=black,
    line width=1.5pt,
]
{1};
\addplot [
    densely dashed,
    domain=-1:1, 
    samples=100, 
    color=blue,
    line width=1pt
    ]
    {1-x^2};
 \addplot [
    domain=-1:1, 
    samples=100, 
    color=orange,
    line width=1.1pt
    ]
    {(1-x^2)^2};
\addplot [
	 densely dashdotted,
    domain=-1:1, 
    samples=100, 
    color=purple,
    line width=1.2pt
    ]
    {(1-x^2)^3};
%\addplot [
%	  densely dotted,
%    domain=-1:1, 
%    samples=100, 
%    color=blue,
%    line width=1.3pt
%    ]
%    {1-abs(x)};
\addplot [
	 loosely dashdotted,
    domain=-3.5:3.5, 
    samples=100, 
    color=red,
    line width=1.4pt
]
{exp(-x^2)};
\addplot [
	  loosely dotted,
    domain=-3.5:3.5, 
    samples=100, 
    color=cyan,
    line width=1.5pt
]
{exp(-x^4)};
\legend{,,Naive,Epanechnikov,Bi-weight,Tri-weight,Gaussian, 4-exponential}
\end{axis}
\end{tikzpicture}
\caption{The shapes of some kernels.}
\label{fig:1}
\end{figure}
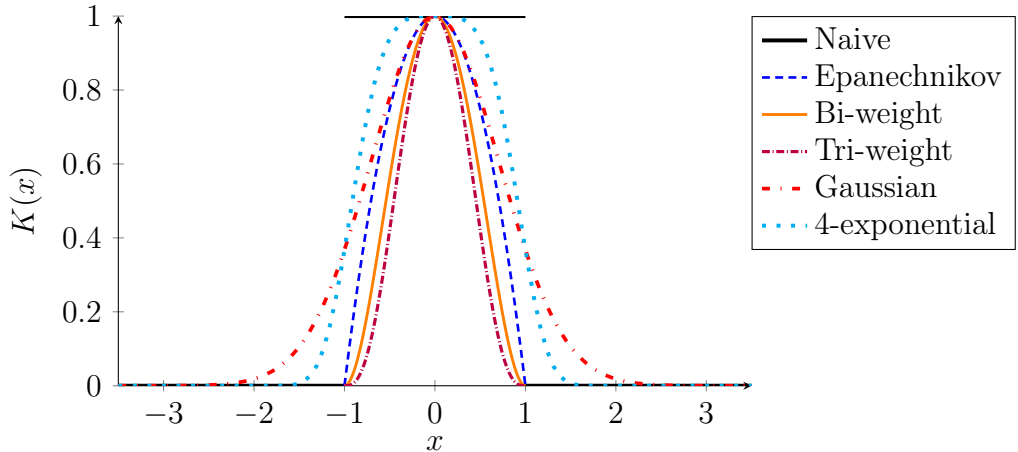

\subsection{Simulated datasets}
In this subsection, we study the performances of our proposed method on the same set of simulated datasets of size $n$ as provided in \cite{cobra}. The input data is either independent and uniformly distributed over $(-1,1)^d$ ({\it uncorrelated} case) or distributed from a Gaussian distribution $\mathcal{N}(0,\Sigma)$ where the covariance matrix $\Sigma$ is defined by $\Sigma_{ij}=2^{-|i-j|}$ for $1\leq i,j\leq d$ ({\it correlated} case). We consider the following models:

\begin{model}{\bf:}
\label{mod:1}
 $n=800, d=50, Y=X_1^2+\exp(-X_2^2).$
\end{model}
\begin{model}{\bf:}
\label{mod:2}
 $n=600, d=100, Y=X_1X_2+X_3^2-X_4X_7+X_8X_{10}-X_6^2+\mathcal{N}(0,0.5).$
\end{model}
\begin{model}{\bf:}
\label{mod:3}
 $n=600, d=100, Y=-\sin(2X_1)+X_2^2+X_3-\exp(-X_4)+\mathcal{N}(0,0.5).$
\end{model}
\begin{model}{\bf:}
\label{mod:4}
 $n=600, d=100, Y=X_1+(2X_2-1)^2+\sin(2\pi X_3)/(2-\sin(2\pi X_3))+\sin(2\pi X_4)+2 \cos(2\pi X_4)+3\sin^2(2\pi X_4)+4\cos^2(2\pi X_4)+\mathcal{N}(0,0.5).$
\end{model}
\begin{model}{\bf:}
\label{mod:5}
 $n=700, d=20, Y=\mathbbm{1}_{\{X_1>0\}}+X_2^3+\mathbbm{1}_{\{X_4+X_6-X_8-X_9>1+X_{14}\}}+\exp(-X_2^2)+\mathcal{N}(0,0.05).$
\end{model}
\begin{model}{\bf:}
\label{mod:6}
 $n=500, d=30, Y=\sum_{k=1}^{10}\mathbbm{1}_{\{X_k<0\}}-\mathbbm{1}_{\{\mathcal{N}(0,1)>1.25\}}.$
\end{model}
\begin{model}{\bf:}
\label{mod:7}
 $n=600, d=300, Y=X_1^2+X_2^2X_3\exp(-|X_4|)+X_6-X_8+\mathcal{N}(0,0.5).$
\end{model}
\begin{model}{\bf:}
\label{mod:8}
 $n=600, d=50, Y=\mathbbm{1}_{\{X_1+X_4^3+X_9+\sin(X_{12}X_{18})+\mathcal{N}(0,0.01)>0.38\}}.$
\end{model}
Moreover, it is interesting to consider some high-dimensional cases as many real problems such as image and signal processing involve these kinds of datasets. Therefore, we also consider the following two high-dimensional models, where the last one is not from \cite{cobra} but a made-up one.
\begin{model}{\bf:}
\label{mod:9}
 $n=500, d=1000, Y=X_1+3X_3^2-2\exp(-X_5)+X_6.$
\end{model}
\begin{model}{\bf:}
\label{mod:10}
 $n=500, d=1500, Y=\exp(X_1)+\exp(-X_1)+\sum_{j=2}^d[\cos(X_j^j))-2\sin(X_j^j)-\exp(-|X_j|)].$
\end{model}
For each model, the proposed method is implemented over $100$ replications. We randomly split $80\%$ of each simulated dataset into two equal parts, $\mathcal{D}_{\ell}$ and $\mathcal{D}_k$ where $\ell=\lceil 0.8\times n/2\rceil-k$, and the remaining $20\%$ will be treated as the corresponding testing data. We measure the performance of any regression method $f$ using {\it mean square error} (MSE) evaluated on the $20\%$-testing data defined by

\begin{equation}
\text{MSE}(f)=\frac{1}{n_{\text{test}}}\sum_{i=1}^{n_{\text{test}}}(y_i^{\text{test}}-f(x_i^{\text{test}}))^2.
\end{equation}

Table~\ref{tab:uncorr} and \ref{tab:corr} below contain the average MSEs and the corresponding standard errors (into brackets) over $100$ runs of {\it uncorrelated} and {\it correlated} cases respectively. In each table, the first block contains five columns corresponding to the following five basic machines ${\bf r}_k=(r_{k,m})_{m=1}^5$:
\begin{itemize}
\item \textcolor{cyan}{\bf Rid}: Ridge regression (R package \texttt{glmnet}, see \cite{glmnet}).
\item \textcolor{cyan}{\bf Las}: Lasso regression (R package \texttt{glmnet}).
\item \textcolor{cyan}{\bf$k$NN}: $k$-nearest neighbors regression (R package \texttt{FNN}, see \cite{FNN}).
\item \textcolor{cyan}{\bf Tr}: Regression tree (R package \texttt{tree}, see \cite{tree}).
\item \textcolor{cyan}{\bf RF}: Random Forest regression (R package \texttt{randomForest}, see \cite{randomForest}).
\end{itemize}
We choose $k=5$ for $k$-NN and $ntree = 300$ for random forest algorithm, and other methods are implemented using the default parameters. The best performance of each method in this block is given in \textbf{boldface}. The second block contains the last seven columns corresponding to the kernel functions used in the combining method where \textcolor{cyan}{\bf COBRA}\footnote{We use the relaxed version of \cite{cobra} with the weights given in equation~(\ref{eq:cobra2}). \texttt{COBRA} library of \texttt{R} programming is used (see, \cite{cobraR}).}, \textcolor{cyan}{\bf Epan}, \textcolor{cyan}{\bf Bi-wgt}, \textcolor{cyan}{\bf Tri-wgt}, \textcolor{cyan}{\bf C-Gaus}, \textcolor{cyan}{\bf Gauss} and \textcolor{cyan}{\bf Exp$4$} respectively stand for classical COBRA, Epanechnikov, Bi-weight, Tri-weight, Compact-support Gaussian, Gaussian and 4-exponential kernels as listed in Table~\ref{tab:kernels}. In this block, the smallest MSE of each case is again written in \textbf{boldface}. For all the compactly supported kernels, we consider $500$ values of $h$ in a uniform grid $\{10^{-100},...,h_{\max}\}$ where $h_{\max}=10$, which is chosen to be large enough, likely to contain the optimal parameter to be searched. For the compactly supported Gaussian, we set $\rho_1=3$ and $\sigma=1$ therefore its support is $[-3,3]$. Lastly, for the two non-compactly supported kernels, Gaussian and 4-exponential, the optimal parameters are estimated using gradient descent algorithm described in the previous section. Note that the results in the first block are not necessarily exactly the same as the ones reported in \cite{cobra} due to the choices of the parameters of the basic machines. 

\begin{figure}[h!]
\includegraphics[width=7cm, height=3.7cm]{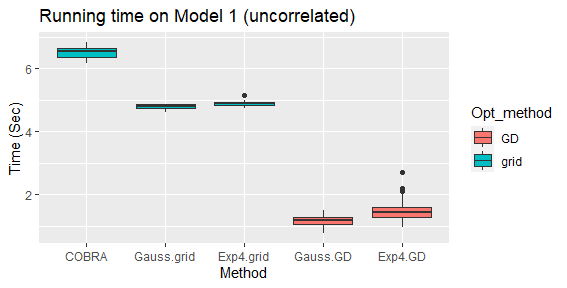}
\includegraphics[width=7cm, height=3.7cm]{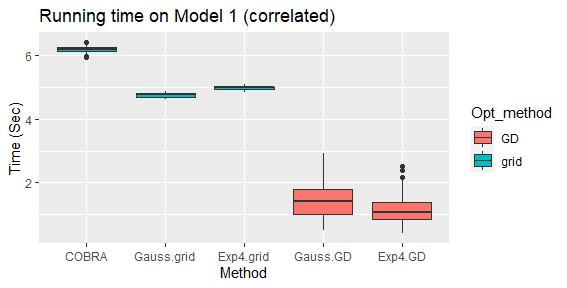}
\includegraphics[width=7cm, height=3.7cm]{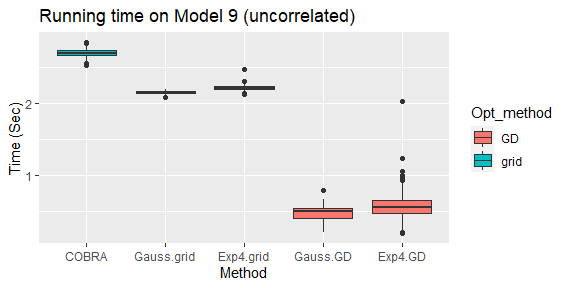}
\includegraphics[width=7cm, height=3.7cm]{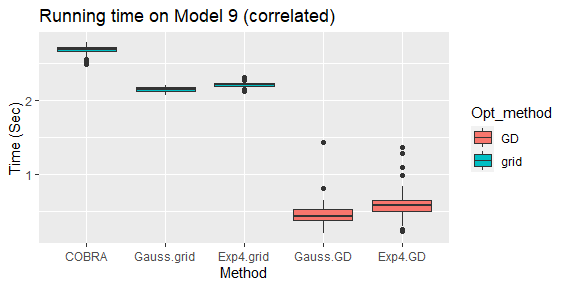}
\caption{Boxplots of runtimes of GD and grid search algorithm implemented on some models.}
\label{fig:2}
\end{figure}

We can easily compare the performances of the combining estimation methods with all the basic machines and among themselves as the results reported in the second block are the straight combinations of those in the first block. In each table, we are interested in comparing the smallest average MSE in the first block to all the columns in the second block. First of all, we can see that all columns of the second block always outperform the best machine of the first block, which illustrates the theoretical result of the combining estimation methods. Secondly, the kernel-based methods beat the first column (classical COBRA) of the second block for almost all kernels. Lastly, the combing estimation method with Gaussian kernel is the absolute winner as the corresponding column is bold in both tables. Note that with the proposed gradient descent algorithm, we can obtain the value of bandwidth parameter with null gradient of cross-validation error defined in equation~(\ref{eq:error2}), which is often better and much faster than the one obtained by the grid search algorithm (2 or 3 times faster). Figure~\ref{fig:2} below contains boxplots of runtimes of 100 runs of Model \ref{mod:1} and \ref{mod:9} of both correlated and uncorrelated cases computed on a machine with the following characteristics:
\begin{itemize}
\item Processor: 2x AMD Opteron 6174, 12C, 2.2GHz, 12x512K L2/12M L3 Cache, 80W ACP, DDR3-1333MHz.
\item Memory: 64GB Memory for 2 CPUs, DDR3, 1333MHz.
\end{itemize}

\begin{sidewaystable}[ph!]
%\footnotesize
\tiny
\centering  % centering table
%\rowcolors{1}{}{lightgray}
\caption{Average MSEs in the uncorrelated case.}
\label{tab:uncorr}

\hspace{0.5em}

\begin{tabular}{ c | c c c c c | c c c c c c c}  % creating 10 columns
\hline\hline                       % inserting double-line 
\textcolor{cyan}{\bf Model} & \textcolor{cyan}{\bf Las} &\textcolor{cyan}{\bf Rid} &\textcolor{cyan}{\bf $k$NN} & \textcolor{cyan}{\bf Tr} & \textcolor{cyan}{\bf RF} & \textcolor{cyan}{\bf COBRA} & \textcolor{cyan}{\bf Epan} & \textcolor{cyan}{\bf Bi-wgt} & \textcolor{cyan}{\bf Tri-wgt}  & \textcolor{cyan}{\bf C-Gaus} & \textcolor{cyan}{\bf Gauss} & \textcolor{cyan}{\bf Exp$4$}\\ [0.5ex]   
\hline  
\multirow{2}{*}{\ref{mod:1}} & \multirow{2}{*}{\raisebox{2ex}{$0.156$}} & \multirow{2}{*}{\raisebox{2ex}{$0.134$}} & \multirow{2}{*}{\raisebox{2ex}{$0.144$}} & \multirow{2}{*}{\raisebox{2ex}{$\bf 0.027$}} & $\multirow{2}{*}{\raisebox{2ex}{0.033}}$ & $0.022$ & $0.020$ & $0.019$ & $0.019$ & $0.019$ & $\textbf{0.018}$ & $0.019$\\
& (0.016) & (0.013)& (0.014) & (0.004) & (0.004) & $(0.004)$ & $(0.003)$ & $(0.003)$ & $(0.003)$ & $(0.003)$ & $(0.002)$ & (0.003)\\
 \hline 
 \multirow{2}{*}{\ref{mod:2}} & \multirow{2}{*}{\raisebox{2ex}{$1.301$}} & \multirow{2}{*}{\raisebox{2ex}{$0.784$}} & \multirow{2}{*}{\raisebox{2ex}{$0.873$}} & \multirow{2}{*}{\raisebox{2ex}{$1.124$}} & $\multirow{2}{*}{\raisebox{2ex}{\textbf{0.707}}}$ & $0.722$ & $0.718$ & $0.712$ & $0.715$ & $0.712$ & $\textbf{0.709}$ & $0.710$\\
 &(0.216) & (0.110)& (0.123) & (0.165) & (0.097) &$(0.065)$ & $(0.079)$ & $(0.080)$ & $(0.079)$ & $(0.079)$ & $(0.078)$ & (0.079)\\
  \hline  
 \multirow{2}{*}{\ref{mod:3}} & \multirow{2}{*}{\raisebox{2ex}{$0.664$}} & \multirow{2}{*}{\raisebox{2ex}{$0.669$}} & \multirow{2}{*}{\raisebox{2ex}{$1.477$}} & \multirow{2}{*}{\raisebox{2ex}{$0.797$}} & $\multirow{2}{*}{\raisebox{2ex}{\textbf{0.629}}}$ & $0.554$ & $ 0.482$ & $0.478$ & $0.476$ & $0.479$ & $\textbf{0.475}$ & $0.483$\\
 &(0.107)& (0.255) & (0.192) & (0.135) & (0.091) & (0.069) & (0.062) & (0.060) & (0.060) & (0.063) & (0.060) & (0.060)\\
   \hline  
 \multirow{2}{*}{\ref{mod:4}} & \multirow{2}{*}{\raisebox{2ex}{$7.783$}} & \multirow{2}{*}{\raisebox{2ex}{$6.550$}} & \multirow{2}{*}{\raisebox{2ex}{$10.238$}} & \multirow{2}{*}{\raisebox{2ex}{$ 3.796$}} & $\multirow{2}{*}{\raisebox{2ex}{\bf 3.774}}$ & $3.608$ & $3.231$ & $3.185$ & $3.153$ & $3.189$ & $\textbf{2.996}$ & $3.186$\\
 & (1.121) & (1.115) & (1.398) & (0.840) & (0.523) & (0.526) & (0.383) & (0.382) & (0.384) & (0.371) & (0.384) & (0.464)\\
   \hline  
 \multirow{2}{*}{\ref{mod:5}} & \multirow{2}{*}{\raisebox{2ex}{$0.508$}} & \multirow{2}{*}{\raisebox{2ex}{$0.518$}} & \multirow{2}{*}{\raisebox{2ex}{$0.699$}} & \multirow{2}{*}{\raisebox{2ex}{$0.575$}} & $\multirow{2}{*}{\raisebox{2ex}{\bf 0.436}}$ & $0.429$ & $0.389$ & $0.387$ & $0.386$ & $0.387$ & $\textbf{0.383}$ & $0.387$\\
 &(0.051) & (0.073)& (0.084) & (0.081) & (0.051) &$(0.035)$ & $(0.031)$ & $(0.030)$ & $(0.030)$ & $(0.030)$ & $(0.030)$ & (0.028)\\
 \hline
 \multirow{2}{*}{\ref{mod:6}} & \multirow{2}{*}{\raisebox{2ex}{$2.693$}} & \multirow{2}{*}{\raisebox{2ex}{$1.958$}} & \multirow{2}{*}{\raisebox{2ex}{$2.675$}} & \multirow{2}{*}{\raisebox{2ex}{$3.065$}} & $\multirow{2}{*}{\raisebox{2ex}{\bf 1.826}}$ & $1.574$ & $1.274$ & $1.259$ & $\bf 1.254$ & $1.270$ & $1.273$ & $1.286$\\
 &(0.537) & (0.292)& (0.349) & (0.475) & (0.262) &$(0.270)$ & $(0.129)$ & $(0.130)$ & $(0.130)$ & $(0.125)$ & $(0.130)$ & (0.130)\\
 \hline
 \multirow{2}{*}{\ref{mod:7}} & \multirow{2}{*}{\raisebox{2ex}{$1.971$}} & \multirow{2}{*}{\raisebox{2ex}{$0.796$}} & \multirow{2}{*}{\raisebox{2ex}{$1.074$}} & \multirow{2}{*}{\raisebox{2ex}{$0.737$}} & $\multirow{2}{*}{\raisebox{2ex}{\bf 0.515}}$ & $0.506$ & $0.472$ & $0.468$ & $0.467$ & $0.469$ & $\textbf{0.451}$ & $0.477$\\
 &(0.410) & (0.132)& (0.152) & (0.109) & (0.073) &$(0.063)$ & $(0.049)$ & $(0.048)$ & $(0.049)$ & $(0.049)$ & $(0.049)$ & (0.067)\\
  \hline   
 \multirow{2}{*}{\ref{mod:8}} & \multirow{2}{*}{\raisebox{2ex}{$0.134$}} & \multirow{2}{*}{\raisebox{2ex}{$0.131$}} & \multirow{2}{*}{\raisebox{2ex}{$0.200$}} & \multirow{2}{*}{\raisebox{2ex}{$0.174$}} & $\multirow{2}{*}{\raisebox{2ex}{\bf 0.127}}$ & $0.104$ & $0.092$ & $\textbf{0.091}$ & $\textbf{0.091}$ & $\textbf{0.091}$ & $\textbf{0.091}$ & $0.094$\\
 &(0.016) & (0.020)& (0.020) & (0.034) & (0.013) &$(0.013)$ & $(0.013)$ & $(0.013)$ & $(0.013)$ & $(0.013)$ & $(0.011)$ & (0.016)\\
  \hline 
 \multirow{2}{*}{\ref{mod:9}} & \multirow{2}{*}{\raisebox{2ex}{$1.592$}} & \multirow{2}{*}{\raisebox{2ex}{$2.948$}} & \multirow{2}{*}{\raisebox{2ex}{$3.489$}} & \multirow{2}{*}{\raisebox{2ex}{$1.830$}} & $\multirow{2}{*}{\raisebox{2ex}{\bf1.488}}$ & $1.130$ & $0.929$ & $0.918$ & $0.914$ & $0.918$ & $\textbf{0.895}$ & $0.993$\\  
 &(0.219) & (0.436)& (0.516) & (0.373) & (0.267) &$(0.151)$ & $(0.128)$ & $(0.127)$ & $(0.130)$ & $(0.124)$ & $(0.126)$ & (0.186)\\
  \hline    
 \multirow{2}{*}{\ref{mod:10}} & \multirow{2}{*}{\raisebox{2ex}{$2012.660$}} & \multirow{2}{*}{\raisebox{2ex}{$\bf 1485.065$}} & \multirow{2}{*}{\raisebox{2ex}{$1778.955$}} & \multirow{2}{*}{\raisebox{2ex}{$3058.381$}} & $\multirow{2}{*}{\raisebox{2ex}{1618.977}}$ & $1511.283$ & $1462.509$ & $1458.306$ & $1459.558$ & $1452.523$ & $\bf1400.365$ & $1414.316$\\
 &(284.391) & (210.816)& (261.396) & (486.504) & (231.555) &$ (129.796)$ & $(143.976)$ & $(142.988)$ & $(142.602)$ & $(141.168)$ & $(143.330)$ & (144.929)\\
 \hline                             % inserts single-line
  \hline  
\end{tabular}  
\caption{Average MSEs in the correlated case.}
\label{tab:corr}

\hspace{0.5em}

\begin{tabular}{ c | c c c c c | c c c c c c c c c}  % creating 10 columns
\hline\hline                       % inserting double-line 
\textcolor{cyan}{\bf Model} & \textcolor{cyan}{\bf Las} &\textcolor{cyan}{\bf Rid} &\textcolor{cyan}{\bf $k$NN} & \textcolor{cyan}{\bf Tr} & \textcolor{cyan}{\bf RF} & \textcolor{cyan}{\bf COBRA} & \textcolor{cyan}{\bf Epan} & \textcolor{cyan}{\bf Bi-wgt} & \textcolor{cyan}{\bf Tri-wgt}  & \textcolor{cyan}{\bf C-Gaus} & \textcolor{cyan}{\bf Gauss} & \textcolor{cyan}{\bf Exp$4$}\\ [0.5ex]   
\hline  
\multirow{2}{*}{\ref{mod:1}} & \multirow{2}{*}{\raisebox{2ex}{$2.294$}} & \multirow{2}{*}{\raisebox{2ex}{$1.947$}} & \multirow{2}{*}{\raisebox{2ex}{$1.941$}} & \multirow{2}{*}{\raisebox{2ex}{$\bf 0.320$}} & $\multirow{2}{*}{\raisebox{2ex}{0.542}}$ & $0.307$ & $0.304$ & $0.301$ & $0.288$ & $0.297$ & $\textbf{0.269}$ & $0.291$\\
 &(0.544 & (0.507)& (0.487) & (0.145) & (0.231) &$(0.129)$ & $(0.105)$ & $(0.111)$ & $(0.103)$ & $(0.104)$ & $(0.092)$ & (0.098)\\
 \hline
 \multirow{2}{*}{\ref{mod:2}} & \multirow{2}{*}{\raisebox{2ex}{$14.273$}} & \multirow{2}{*}{\raisebox{2ex}{$8.442$}} & \multirow{2}{*}{\raisebox{2ex}{$8.572$}} & \multirow{2}{*}{\raisebox{2ex}{$ 6.796$}} & $\multirow{2}{*}{\raisebox{2ex}{\textbf{5.135}}}$ & $5.345$ & $4.582$ & $4.529$ & $4.491$ & $4.541$ & $\textbf{4.377}$ & $4.910$\\
 &(2.593) & (1.912)& (1.751) & (1.548) & (1.372) &$(1.194)$ & $(0.941)$ & $(0.934)$ & $(0.922)$ & $(0.896)$ & $(0.905)$ & (1.181)\\
  \hline
 \multirow{2}{*}{\ref{mod:3}} & \multirow{2}{*}{\raisebox{2ex}{$7.996$}} & \multirow{2}{*}{\raisebox{2ex}{$6.266$}} & \multirow{2}{*}{\raisebox{2ex}{$8.704$}} & \multirow{2}{*}{\raisebox{2ex}{$4.110$}} & $\multirow{2}{*}{\raisebox{2ex}{\bf 3.722}}$ & $3.327$ & $2.598$ & $2.536$ & $2.444$ & $2.554$ & $\textbf{2.168}$ & $2.357$\\
 &(3.393) & (3.296)& (3.523) & (2.894) & (2.956) & $(1.006)$ & $(0.912)$ & $(0.944)$ & $(0.840)$ & $(0.907)$ & $(0.680)$ & (0.756) \\
   \hline
 \multirow{2}{*}{\ref{mod:4}} & \multirow{2}{*}{\raisebox{2ex}{$61.474$}} & \multirow{2}{*}{\raisebox{2ex}{$42.351$}} & \multirow{2}{*}{\raisebox{2ex}{$46.934$}} & \multirow{2}{*}{\raisebox{2ex}{$\bf 8.855$}} & $\multirow{2}{*}{\raisebox{2ex}{13.381}}$ & $9.599$ & $10.511$ & $9.963$ & $9.682$ & $10.085$ & $\textbf{9.056}$ & $9.713$\\
 &(13.986) & (11.622)& (12.543) & (3.480) & (5.549) &$(4.125)$ & $(2.961)$ & $(3.101)$ & $(2.860)$ & $(2.904)$ & $(2.407)$ & (2.695) \\
   \hline
 \multirow{2}{*}{\ref{mod:5}} & \multirow{2}{*}{\raisebox{2ex}{$6.805$}} & \multirow{2}{*}{\raisebox{2ex}{$7.479$}} & \multirow{2}{*}{\raisebox{2ex}{$10.342$}} & \multirow{2}{*}{\raisebox{2ex}{$\bf 4.000$}} & $\multirow{2}{*}{\raisebox{2ex}{4.880}}$ & $3.225$ & $2.640$ & $2.401$ & $2.235$ & $2.412$ & $\textbf{1.792}$ & $2.194$\\
 &(3.685) & (5.336)& (5.425) & (3.144) & (3.787) &$(2.088)$ & $(1.455)$ & $(1.387)$ & $(1.250 )$ & $(1.355)$ & $(0.913)$ & (1.242)\\
 \hline   
 \multirow{2}{*}{\ref{mod:6}} & \multirow{2}{*}{\raisebox{2ex}{$4.221$}} & \multirow{2}{*}{\raisebox{2ex}{$2.087$}} & \multirow{2}{*}{\raisebox{2ex}{$4.461$}} & \multirow{2}{*}{\raisebox{2ex}{$3.408$}} & $\multirow{2}{*}{\raisebox{2ex}{\bf 1.701}}$ & $1.493$ & $1.271$ & $1.238$ & $1.217$ & $1.248$ & $\textbf{1.097 }$ & $1.270$\\
 &(0.848) & (0.485)& (0.599) & (0.636) & (0.288) &$(0.326)$ & $(0.149)$ & $(0.146)$ & $(0.143)$ & $(0.148)$ & $(0.145)$ & (0.386)\\
 \hline   
 \multirow{2}{*}{\ref{mod:7}} & \multirow{2}{*}{\raisebox{2ex}{$17.875$}} & \multirow{2}{*}{\raisebox{2ex}{$4.695$}} & \multirow{2}{*}{\raisebox{2ex}{$5.591$}} & \multirow{2}{*}{\raisebox{2ex}{$4.132$}} & $\multirow{2}{*}{\raisebox{2ex}{\bf 3.081}}$ & $3.304$ & $2.819$ & $2.779$ & $2.736$ & $2.788$ & $\textbf{2.640}$ & $ 2.979$\\
 &(5.632) & (1.318)& (1.418) & (1.360) & (1.091) &$(0.799$ & $(0.636)$ & $(0.614)$ & $(0.605)$ & $(0.623)$ & $(0.590)$ & (0.764)\\
  \hline    
 \multirow{2}{*}{\ref{mod:8}} & \multirow{2}{*}{\raisebox{2ex}{$0.139$}} & \multirow{2}{*}{\raisebox{2ex}{$0.133$}} & \multirow{2}{*}{\raisebox{2ex}{$ 0.201$}} & \multirow{2}{*}{\raisebox{2ex}{$0.159$}} & $\multirow{2}{*}{\raisebox{2ex}{\bf 0.121}}$ & $0.102$ & $0.100$ & $0.100$ & $0.100$ & $0.100$ & $\textbf{0.092}$ & $\bf0.092$\\
 &(0.016) & (0.020)& (0.019) & (0.035) & (0.013) &$(0.021)$ & $(0.020)$ & $(0.021)$ & $(0.020)$ & $(0.020)$ & $(0.021)$ & (0.018) \\
  \hline
 \multirow{2}{*}{\ref{mod:9}} & \multirow{2}{*}{\raisebox{2ex}{$43.445$}} & \multirow{2}{*}{\raisebox{2ex}{$37.827$}} & \multirow{2}{*}{\raisebox{2ex}{$43.991$}} & \multirow{2}{*}{\raisebox{2ex}{$\bf15.258$}} & $\multirow{2}{*}{\raisebox{2ex}{16.957}}$ & $13.505$ & $11.303$ & $11.007$ & $11.067$ & $11.206$ & $\textbf{10.303}$ & $12.346$\\
 &(12.210) & (12.201)& (12.920) & (8.119) & (8.774) &$(4.822)$ & $(3.891)$ & $(3.815)$ & $(3.949)$ & $(3.960)$ & $(3.634)$ & (5.014)\\
 %\hline                          % inserts single-line
 \hline  
\multirow{2}{*}{\ref{mod:10}} & \multirow{2}{*}{\raisebox{2ex}{$7235.062$}} & \multirow{2}{*}{\raisebox{2ex}{$\bf 5244.843$}} & \multirow{2}{*}{\raisebox{2ex}{$7636.811$}} & \multirow{2}{*}{\raisebox{2ex}{$13014.596$}} & $\multirow{2}{*}{\raisebox{2ex}{7092.741}}$ & $5147.950$ & $4717.225$ & $4669.516$ & $4663.430$ & $4697.019$ & $\textbf{4660.043}$ & $5073.591$\\
 &(1100.579) & (996.181)& (1159.445) & (2020.133) & (1030.249) &$(835.384)$ & $(703.049)$ & $(696.027)$ & $(687.474)$ & $(681.370)$ & $(764.363)$ & (1022.894)\\
  \hline     
   \hline      
\end{tabular}
\end{sidewaystable}

\subsection{Real public datasets}
In this part, we consider three public datasets which are available and easily accessible on the internet. The first dataset ({\bf Abalone}, available at \cite{abaloneData}) contains $4177$ rows and $9$ columns of measurements of abalones observed in Tasmania, Australia. We are interested in predicting the age of each abalone through the number of rings using its physical characteristics such as gender, size, weight,  etc. The second dataset ({\bf House}, available at \cite{HouseKC2016}) comprises house sale prices for King County including Seattle. It contains homes sold between May 2014 and May 2015. The dataset consists of $21613$ rows of houses and $21$ columns of characteristics of each house including ID, Year of sale, Size, Location, etc. In this case, we want to predict the price of each house using all of its quantitative characteristics. 

Notice that Model~\ref{mod:6} and \ref{mod:8} of the previous subsection are about predicting integer labels of the response variable. Analogously, the last dataset ({\bf Wine}, see \cite{redWineData,wineArticle}), which was also considered in \cite{cobra}, containing $1599$ rows of different types of wines and $12$ columns corresponding to different substances of red wines including the amount of different types of acids, sugar, chlorides, PH, etc. The variable of interest is {\it quality} which scales from $3$ to $8$ where $8$ represents the best quality. We aim at predicting the quality of each wine, which is treated as a continuous variable, using all of its substances.

The five primary machines are Ridge, LASSO, $k$NN, Tree and Random Forest regression. In this case, the parameter $ntree=500$ for random forest, and $k$NN is implemented using $k=20,12$ and $5$ for Abalone, House and Wine dataset respectively. The five machines are combined using the classical method by \cite{cobra} and the kernel-based method with Gaussian kernel as it is the most outstanding one among all the kernel functions. In this case, the search for parameter $h$ for the classical COBRA method is performed using a grid of size $300$. In addition, due to the scaling issue, we measure the performance of any method $f$ in this case using average \textit{root mean square error} (RMSE) defined by,

\begin{equation}
\text{RMSE}(f)=\sqrt{\text{MSE}(f)}=\sqrt{\frac{1}{n_{\text{test}}}\sum_{i=1}^{n_{\text{test}}}(y_i^{\text{test}}-f(x_i^{\text{test}}))^2}.
\end{equation}
The average RMSEs obtained from $100$ independent runs, evaluated on $20\%$-testing data of the three public datasets, are provided in Table~\ref{tab:realdata} below (the first three rows). We observe that random forest is the best estimator among all the basic machines in the first block, and the proposed method either outperforms other columns ({\bf Wine} and {\bf Abalone}) or biases towards the best basic machine ({\bf House}). Moreover, the performances of kernel-based method always exceed the ones of the classical method by \cite{cobra}.

\subsection{Real private datasets}
The results presented in this subsection are obtained from two private datasets. The first dataset contains six columns corresponding to the six variables including \textit{Air temperature, Input Pressure, Output Pressure, Flow, Water Temperature} and \textit{Power Consumption} along with $2026$ rows of hourly observations of these measurements of an air compressor machine provided by \cite{CHM}. The goal is to predict the power consumption of this machine using the five remaining explanatory variables. The second dataset is provided by the wind energy company Ma$\ddot{\text{\i}}$a Eolis. It contains 8721 observations of seven variables representing 10-minute measurements of \textit{Electrical power, Wind speed, Wind direction, Temperature, Variance of wind speed} and \textit{Variance of wind direction} measured from a wind turbine of the company (see, \cite{wind}). In this case, we aim at predicting the electrical power produced by the turbine using the remaining six measurements as explanatory variables. We use the same set of parameters as in the previous subsection except for $k$NN where in this case $k=10$ and $k=7$ are used for air compressor and wind turbine dataset respectively. The results obtained from $100$ independent runs of the methods are presented in the last two rows (\textbf{Air} and \textbf{Turbine}) of Table~\ref{tab:realdata} below. We observe on one hand that the proposed method (\textcolor{cyan}{\bf Gauss}) outperforms both the best basic machines (\textcolor{cyan}{\bf RF}) and the classical method by \cite{cobra} in the case of {\bf Turbine} dataset. On the other hand, the performance of our method approaches the performance of the best basic machine (\textcolor{cyan}{\bf Las}) and outperforms the classical COBRA in the case of \textbf{Air} dataset. Moreover, boxplots of runtimes (100 runs) measured on {\bf Wine} and {\bf Turbine} datasets (computed using the same machine as described in the subsection of simulated data) are also given in Figure~\ref{fig:3} below.

\begin{table}[ph!]
\centering
\tiny
\caption{Average RMSEs of real datasets.}
\label{tab:realdata}
\begin{tabular}{l c c c c c | c c}  % creating 10 columns
\hline\hline                       % inserting double-line 
\textcolor{cyan}{\bf Model} & \textcolor{cyan}{\bf Las} &\textcolor{cyan}{\bf Rid} &\textcolor{cyan}{\bf $k$NN} & \textcolor{cyan}{\bf Tr} & \textcolor{cyan}{\bf RF} & \textcolor{cyan}{\bf COBRA} & \textcolor{cyan}{\bf Gauss}\\ [0.5ex] 
\hline  
\multirow{2}{*}{\bf House} & \multirow{2}{*}{\raisebox{2ex}{241083.959}} & \multirow{2}{*}{\raisebox{2ex}{241072.974}} & \multirow{2}{*}{\raisebox{2ex}{245153.608}} & \multirow{2}{*}{\raisebox{2ex}{254099.652}} & \multirow{2}{*}{\raisebox{2ex}{$\bf 205943.768$}} & $\multirow{2}{*}{\raisebox{2ex}{223596.317}}$& $\textbf{209955.276}$ \\
& (8883.107) & (8906.332)& (23548.367) & (9350.885) & (7496.766) &$(13299.934)$ & $(7815.623)$\\
\hline
\multirow{2}{*}{\bf Wine} & \multirow{2}{*}{\raisebox{2ex}{0.0.660}} & \multirow{2}{*}{\raisebox{2ex}{0.685}} & \multirow{2}{*}{\raisebox{2ex}{0.767}} & \multirow{2}{*}{\raisebox{2ex}{0.711}} & \multirow{2}{*}{\raisebox{2ex}{$\bf 0.623$}} & $0.650$ & $\multirow{2}{*}{\raisebox{2ex}{\textbf{0.617}}}$\\
& (0.029) & (0.053)& (0.031) & (0.030) & (0.028) &$(0.026)$ & $(0.020)$\\
 \hline
  \multirow{2}{*}{\bf Abalone} & \multirow{2}{*}{\raisebox{2ex}{2.204}} & \multirow{2}{*}{\raisebox{2ex}{2.215}} & \multirow{2}{*}{\raisebox{2ex}{2.175}} & \multirow{2}{*}{\raisebox{2ex}{2.397}} & \multirow{2}{*}{\raisebox{2ex}{$\bf 2.153$}} & $\multirow{2}{*}{\raisebox{2ex}{2.171}}$ & $\textbf{2.128}$\\
& (0.071) & (0.075)& (0.062) & (0.072) & (0.060) &$(0.081)$ & $(0.057)$\\
\hline
\multirow{2}{*}{\bf Air} & \multirow{2}{*}{\raisebox{2ex}{\bf163.099}} & \multirow{2}{*}{\raisebox{2ex}{164.230}} & \multirow{2}{*}{\raisebox{2ex}{241.657}} & \multirow{2}{*}{\raisebox{2ex}{351.317}} & \multirow{2}{*}{\raisebox{2ex}{$174.836$}} & $\multirow{2}{*}{\raisebox{2ex}{172.858}}$ & \textbf{163.253}\\
& (3.694) & (3.746)& (5.867) & (31.876) & (6.554) &$(7.644)$ & $(3.333)$\\
 \hline
 \multirow{2}{*}{\bf Turbine} & \multirow{2}{*}{\raisebox{2ex}{70.051}} & \multirow{2}{*}{\raisebox{2ex}{68.987}} & \multirow{2}{*}{\raisebox{2ex}{44.516}} & \multirow{2}{*}{\raisebox{2ex}{81.714}} & \multirow{2}{*}{\raisebox{2ex}{$\bf 38.894$}} & $\multirow{2}{*}{\raisebox{2ex}{38.927}}$ & \textbf{37.135}\\
& (4.986) & (3.413)& (1.671) & (4.976) & (1.506) &$(1.561)$ & $(1.555)$\\
 \hline
 \hline
\end{tabular}
\end{table}

\begin{figure}[h!]
\includegraphics[width=7cm, height=3.7cm]{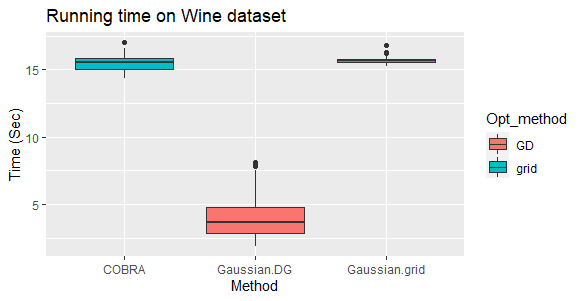}
\includegraphics[width=7cm, height=3.7cm]{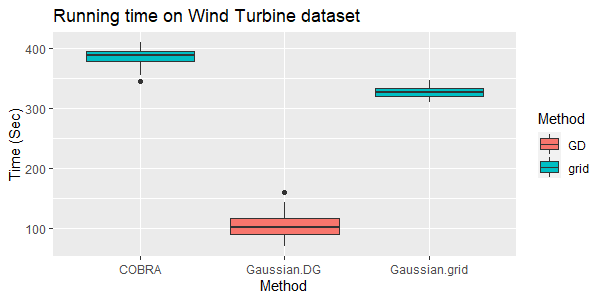}
\caption{Boxplots of runtimes of GD and grid search algorithm implemented on {\bf Wine} and {\bf Turbine} datasets.}
\label{fig:3}
\end{figure}
\section{Conclusion}

In this study, we investigate and extend the context of a naive kernel-based consensual regression aggregation method by \cite{cobra} to a more general regular kernel-based framework. Moreover, an optimization algorithm based on gradient descent is proposed to efficiently and rapidly estimate the key parameter of the method. It is also shown through several numerical simulations that the performance of the method is improved significantly with smoother kernel functions.

%\noindent In the future work, it could be very interesting to study the performance of the proposed method on high dimensional features due to its consistence inheritance property. The combination of high dimensional features defined by the predictions of the basic machines and dimensional reduction techniques such as random projection could be considered. It would be very interesting to investigate the trade-off between information brought by high-dimensional features of predictions and the reduced dimension suggested by the dimensional reduction strategy.
\noindent In practice, the performance of the consensual aggregation depends both on the performance of the individual regression machines, and on the final combination here involving kernel functions. Since the calibration of hyperparameters may be critical in both steps, it could be very interesting to investigate in future work how automated machine learning models can improve the performances of the global model.

\section{Proofs}
\label{sec:proof}

%==================================== Lemma 1 ========================================%

The following lemma, which is a variant of lemma 4.1 in \cite{bookDistributionFree} related to the property of binomial random variables, is needed. 

\begin{lemma}%(\cite{bookDistributionFree})
\label{lem:1}
Let $B(n,p)$ be the binomial random variable with parameters $n$ and $p$. Then
\begin{enumerate}
\item\label{itm:1} For any $c>0$,
 \begin{align*}
	\mathbb{E}\Big[\frac{1}{c+B(n,p)}\Big]\leq \frac{2}{p(n+1)}.
\end{align*}
\item\label{itm:2} \begin{align*}
	\mathbb{E}\Big[\frac{1}{B(n,p)}\mathbbm{1}_{B(n,p)>0}\Big]\leq \frac{2}{p(n+1)}.
\end{align*}
\end{enumerate}
\end{lemma}

\begin{prooflemma}
\begin{enumerate}
\item For any $c>0$, one has
\begin{align*}
	\mathbb{E}\Big[\frac{1}{c+B(n,p)}\Big]&=\sum_{k=0}^n\frac{1}{c+k}\times\frac{n!}{(n-k)!k!}p^k(1-p)^{n-k}\\
	&= \sum_{k=0}^n\frac{1}{k+1}\times\frac{k+1}{k+c}\times\frac{n!}{(n-k)!k!}p^k(1-p)^{n-k}\\
	&\leq\frac{2}{p(n+1)} \sum_{k=0}^{n}\frac{(n+1)!p^{k+1}(1-p)^{n+1-(k+1)}}{[n+1-(k+1)]!(k+1)!}\\
	&\leq\frac{2}{p(n+1)} \sum_{k=0}^{n+1}\frac{(n+1)!p^k(1-p)^{n+1-k}}{[n+1-k]!k!}\\
	&=\frac{2}{p(n+1)} (p+1-p)^{n+1}\\
	&=\frac{2}{p(n+1)}
\end{align*}
\item 
\begin{align*}
	\mathbb{E}\Big[\frac{1}{B(n,p)}\mathbbm{1}_{B(n,p)>0}\Big]&\leq \mathbb{E}\Big[\frac{2}{1+B(n,p)}\Big]\\
	&= \sum_{k=0}^n\frac{2}{k+1}\times\frac{n!}{(n-k)!k!}p^k(1-p)^{n-k}\\
	&=\frac{2}{p(n+1)} \sum_{k=0}^{n}\frac{(n+1)!p^{k+1}(1-p)^{n+1-(k+1)}}{[n+1-(k+1)]!(k+1)!}\\
	&\leq\frac{2}{p(n+1)} \sum_{k=0}^{n+1}\frac{(n+1)!p^k(1-p)^{n+1-k}}{[n+1-k]!k!}\\
	&=\frac{2}{p(n+1)} (p+1-p)^{n+1}\\
	&=\frac{2}{p(n+1)}
\end{align*}
\end{enumerate}
\QED
\end{prooflemma}

%==================================== Proof of Proposition 1 ========================================%

\begin{proofprop}
For any square integrable function with respect to $\textbf{r}_k(X)$, one has

\begin{align*}
\mathbb{E}\Big[|g_n(\textbf{r}_k(X))-g^*(X)|^2\Big]&= \mathbb{E}\Big[|g_n(\textbf{r}_k(X))-g^*(\textbf{r}_k(X))+g^*(\textbf{r}_k(X))-g^*(X)|^2\Big]\\
&=\mathbb{E}\Big[|g_n(\textbf{r}_k(X))-g^*(\textbf{r}_k(X))|^2\Big]\\
&\quad+2\mathbb{E}\Big[(g_n(\textbf{r}_k(X))-g^*(\textbf{r}_k(X)))(g^*(\textbf{r}_k(X))-g^*(X))\Big]\\
&\quad+\mathbb{E}\Big[|g^*(\textbf{r}_k(X))-g^*(X)|^2\Big].
\end{align*}
We consider the second term of the right hand side of the last equality,
\begin{align*}
{}&\ \mathbb{E}\Big[(g_n(\textbf{r}_k(X))-g^*(\textbf{r}_k(X)))(g^*(\textbf{r}_k(X))-g^*(X))\Big]\\
=&\ \mathbb{E}_{\textbf{r}_k(X)}\Big[\mathbb{E}_{X}\Big[(g_n(\textbf{r}_k(X))-g^*(\textbf{r}_k(X)))(g^*(\textbf{r}_k(X))-g^*(X))\Big|\textbf{r}_k(X)\Big]\Big]\\
=&\ \mathbb{E}_{\textbf{r}_k(X)}\Big[(g_n(\textbf{r}_k(X))-g^*(\textbf{r}_k(X)))(g^*(\textbf{r}_k(X))-\mathbb{E}[g^*(X)|\textbf{r}_k(X)])\Big]\\
=&\ 0
\end{align*}
where $g^*(\textbf{r}_k(X))=\mathbb{E}[g^*(X)|\textbf{r}_k(X)]$ thanks to the definition of $g^*(\textbf{r}_k(X))$ and the tower property of conditional expectation. It remains to check that
 $$\mathbb{E}\Big[|g^*(\textbf{r}_k(X))-g^*(X)|^2\Big]\leq \inf_{f\in\mathcal{G}}\mathbb{E}\Big[|f(\textbf{r}_k(X))-g^*(X)|^2\Big].$$
For any function $f$ s.t $\mathbb{E}\Big[|f(\textbf{r}_k(X))|^2\Big]<+\infty$, one has
 \begin{align*}
 \mathbb{E}\Big[|f(\textbf{r}_k(X))-g^*(X)|^2\Big]&= \mathbb{E}\Big[|f(\textbf{r}_k(X))-g^*(\textbf{r}_k(X))+g^*(\textbf{r}_k(X))-g^*(X)|^2\Big]\\
&= \mathbb{E}\Big[|f(\textbf{r}_k(X))-g^*(\textbf{r}_k(X))|^2\Big]\\
 &\quad+2\mathbb{E}\Big[(f(\textbf{r}_k(X))-g^*(\textbf{r}_k(X)))(g^*(\textbf{r}_k(X))-g^*(X))\Big]\\
&\quad+\mathbb{E}\Big[|g^*(\textbf{r}_k(X))-g^*(X)|^2\Big].
\end{align*}
Similarly,
\begin{align*}
\mathbb{E}\Big[(f(\textbf{r}_k(X))-g^*(\textbf{r}_k(X)))(g^*(\textbf{r}_k(X))-g^*(X))\Big] = 0.
\end{align*}
Therefore,
\begin{align*}
 \mathbb{E}\Big[|f(\textbf{r}_k(X))-g^*(X)|^2\Big]&= \mathbb{E}\Big[|f(\textbf{r}_k(X))-g^*(\textbf{r}_k(X))|^2\Big]\\
 &\quad+\mathbb{E}\Big[|g^*(\textbf{r}_k(X))-g^*(X)|^2\Big].
\end{align*}
 As the first term of the right-hand side is nonnegative thus,
 $$\mathbb{E}\Big[|g^*(\textbf{r}_k(X))-g^*(X)|^2\Big]\leq \inf_{f\in\mathcal{G}}\mathbb{E}\Big[|f(\textbf{r}_k(X))-g^*(X)|^2\Big].$$
 Finally, we can conclude that
 \begin{align*}
	\mathbb{E}\Big[|g_n(\textbf{r}_k(X))-g^*(X)|^2\Big]&\leq \mathbb{E}\Big[|g_n(\textbf{r}_k(X))-g^*(\textbf{r}_k(X))|^2\Big]\\
	&\quad+\inf_{f\in\mathcal{G}}\mathbb{E}\Big[|f(\textbf{r}_k(X))-g^*(X)|^2\Big].
\end{align*}
\end{proofprop}
We obtain the particular case by restricting $\mathcal{G}$ to be the coordinates of $\textbf{r}_k$, one has
 \begin{align*}
	\mathbb{E}\Big[|g_n(\textbf{r}_k(X))-g^*(X)|^2\Big]&\leq \mathbb{E}\Big[|g_n(\textbf{r}_k(X))-g^*(\textbf{r}_k(X))|^2\Big]\\
	&\quad+\min_{1\leq m\leq M}\mathbb{E}\Big[|r_{k,m}(X)-g^*(X)|^2\Big].
\end{align*}
\QED

%In order to prove the \textbf{Proposition~2}, we need the following technical lemma.
%\begin{lemma}%(\cite{bookDistributionFree})
%\label{lem:1}
%Let $B(n,p)$ be the binomial random variable with parameters $n$ and $p$. Then
%\begin{enumerate}
%\item\label{itm:1} For any $c>0$,
% \begin{align*}
%	\mathbb{E}\Big[\frac{1}{c+B(n,p)}\Big]\leq \frac{2}{p(n+1)}.
%\end{align*}
%\item\label{itm:2} \begin{align*}
%	\mathbb{E}\Big[\frac{1}{B(n,p)}\mathbbm{1}_{B(n,p)>0}\Big]\leq \frac{2}{p(n+1)}.
%\end{align*}
%\end{enumerate}
%\end{lemma}
%%%%%%%%%%%%%==================== Proof of Proposition 2 ==================%%%%%%%%%%%%%%
\begin{proofprop}
The procedure of proving this result is indeed the procedure of checking the conditions of Stone's theorem (see, for example, \cite{stone1977} and Chapter 4 of \cite{bookDistributionFree}) which is also used in the classical method by \cite{cobra}. First of all, using the inequality: $(a+b+c)^2\leq 3(a^2+b^2+c^2)$, one has
\begin{align*}
\ \mathbb{E}\Big[|g_n(\textbf{r}_k(X))-g^*(\textbf{r}_k(X))|^2\Big]&=\mathbb{E}\Big[\Big|\sum_{i=1}^{\ell}W_{n,i}(X)Y_i-g^*(\textbf{r}_k(X))\Big|^2\Big]\\
&=\mathbb{E}\Big[\Big|\sum_{i=1}^{\ell}W_{n,i}(X)[Y_i-g^*(\textbf{r}_k(X_i))]\\
&\quad+\sum_{i=1}^{\ell}W_{n,i}(X)[g^*(\textbf{r}_k(X_i))-g^*(\textbf{r}_k(X))]\\
&\quad+\sum_{i=1}^{\ell}W_{n,i}(X)g^*(\textbf{r}_k(X))-g^*(\textbf{r}_k(X))\Big|^2\Big]\\
&\leq 3\mathbb{E}\Big[\Big|\sum_{i=1}^{\ell}W_{n,i}(X)[g^*(\textbf{r}_k(X_i))-g^*(\textbf{r}_k(X))]\Big|^2\Big]\\
&\quad+3\mathbb{E}\Big[\Big|\sum_{i=1}^{\ell}W_{n,i}(X)[Y_i-g^*(\textbf{r}_k(X_i))]\Big|^2\Big]\\
&\quad+3\mathbb{E}\Big[\Big|g^*(\textbf{r}_k(X))\sum_{i=1}^{\ell}(W_{n,i}(X)-1)\Big|^2\Big].
\end{align*}
The three terms of the right-hand side are denoted by $A.1, A.2$ and $A.3$ respectively, thus one has
\begin{align*}
\mathbb{E}\Big[|g_n(\textbf{r}_k(X))-g^*(\textbf{r}_k(X))|^2\Big]&\leq 3(A.1+A.2+A.3).
\end{align*}
To prove the result, it is enough to prove that the three terms \textbf{$A.1,A.2$} and \textbf{$A.3$} vanish under the assumptions of \textbf{Proposition~2}. We deal with the first term \textbf{$A.1$} in the following proposition.

\begin{propA}
Under the assumptions of \textbf{Proposition~2},
\begin{align*}
\lim_{\ell\rightarrow+\infty}\mathbb{E}\Big[\Big|\sum_{i=1}^{\ell}W_{n,i}(X)[g^*(\textbf{r}_k(X_i))-g^*(\textbf{r}_k(X))]\Big|^2\Big]=0.
\end{align*}
\end{propA}
\begin{proofA}
Using Cauchy-Schwarz's inequality, one has
\begin{align*}
A.1&=\mathbb{E}\Big[\Big|\sum_{i=1}^{\ell}W_{n,i}(X)[g^*(\textbf{r}_k(X_i))-g^*(\textbf{r}_k(X))]\Big|^2\Big]\\
&=\mathbb{E}\Big[\Big|\sum_{i=1}^{\ell}\sqrt{W_{n,i}(X)}\sqrt{W_{n,i}(X)}[g^*(\textbf{r}_k(X_i))-g^*(\textbf{r}_k(X))]\Big|^2\Big]\\
{}&\leq \mathbb{E}\Big[\Big(\sum_{i=1}^{\ell}W_{n,i}(X)\Big)\sum_{i=1}^{\ell}W_{n,i}(X)[g^*(\textbf{r}_k(X_i))-g^*(\textbf{r}_k(X))]^2\Big]\\
&=\mathbb{E}\Big[\sum_{i=1}^{\ell}W_{n,i}(X)[g^*(\textbf{r}_k(X_i))-g^*(\textbf{r}_k(X))]^2\Big]\\
&\eqdef A_n.
\end{align*}
Note that the regression function $g^*$ satisfies $\mathbb{E}[|g^*(\textbf{r}_k(X))|^2]<+\infty$, thus it can be approximated in $L_2$ sense by a continuous function with compact support named $\tilde{g}$ (see, for example, Theorem A.1 in \cite{bookProbabTheoryPattern}). This means that for any $\varepsilon>0$, there exists a continuous function with compact support $\tilde{g}$ such that,
$$\mathbb{E}[|g^*(\textbf{r}_k(X))-\tilde{g}(\textbf{r}_k(X))|^2]<\varepsilon.$$
Thus, one has
\begin{align*}
A_n&=\mathbb{E}\Big[\sum_{i=1}^{\ell}W_{n,i}(X)[g^*(\textbf{r}_k(X_i))-g^*(\textbf{r}_k(X))]^2\Big]\\
&\leq 3\mathbb{E}\Big[\sum_{i=1}^{\ell}W_{n,i}(X)[g^*(\textbf{r}_k(X_i))-\tilde{g}(\textbf{r}_k(X_i))]^2\Big]\\
&\quad+3\mathbb{E}\Big[\sum_{i=1}^{\ell}W_{n,i}(X)[\tilde{g}(\textbf{r}_k(X_i))-\tilde{g}(\textbf{r}_k(X))]^2\Big]\\
&\quad+3\mathbb{E}\Big[\sum_{i=1}^{\ell}W_{n,i}(X)[\tilde{g}(\textbf{r}_k(X))-g^*(\textbf{r}_k(X))]^2\Big]\\
&\eqdef 3(A_{n1}+A_{n2}+A_{n3}).
\end{align*}
We deal with each term of the last upper bound as follows.

% ============= An1 ===============
\begin{itemize}
\item Computation of \textbf{$A_{n3}$}: applying the definition of $\tilde{g}$,
\begin{align*}
A_{n3}&=\mathbb{E}\Big[\sum_{i=1}^{\ell}W_{n,i}(X)[\tilde{g}(\textbf{r}_k(X))-g^*(\textbf{r}_k(X))]^2\Big]\\
&\leq \mathbb{E}\Big[|\tilde{g}(\textbf{r}_k(X))-g^*(\textbf{r}_k(X))|^2\Big]< \varepsilon.
\end{align*}
\item Computation of \textbf{$A_{n1}$}: denoted by $\mu$ the distribution of $X$. Thus,
\begin{align*}
A_{n1}&=\mathbb{E}\Big[\sum_{i=1}^{\ell}W_{n,i}(X)|g^*(\textbf{r}_k(X_i))-\tilde{g}(\textbf{r}_k(X_i))|^2\Big]\\
&=\ell\mathbb{E}\Big[W_{n,1}(X)|g^*(\textbf{r}_k(X_1))-\tilde{g}(\textbf{r}_k(X_1))|^2\Big]\\
&=\ell\mathbb{E}\Big[\frac{K_h(\textbf{r}_k(X)-\textbf{r}_k(X_1))}{\sum_{j=1}^{\ell}K_h(\textbf{r}_k(X)-\textbf{r}_k(X_j))}|g^*(\textbf{r}_k(X_1))-\tilde{g}(\textbf{r}_k(X_1))|^2\Big]\\
&=\ell\mathbb{E}_{\mathcal{D}_k}\Big[\mathbb{E}_{\{X_j\}_{j=1}^{\ell}}\Big[\int\frac{K_h(\textbf{r}_k(v)-\textbf{r}_k(X_1))}{\sum_{j=1}^{\ell}K_h(\textbf{r}_k(v)-\textbf{r}_k(X_j))}\times\\
&\quad\ |g^*(\textbf{r}_k(X_1))-\tilde{g}(\textbf{r}_k(X_1))|^2\mu(dv)\Big|\mathcal{D}_k\Big]\Big]\\
&=\ell\mathbb{E}_{\mathcal{D}_k}\Big[\mathbb{E}_{\{X_j\}_{j=2}^{\ell}}\Big[\int\int|g^*(\textbf{r}_k(u))-\tilde{g}(\textbf{r}_k(u))|^2\times\\
&\quad\frac{K_h(\textbf{r}_k(v)-\textbf{r}_k(u))}{K_h(\textbf{r}_k(v)-\textbf{r}_k(u))+\sum_{j=2}^{\ell}K_h(\textbf{r}_k(v)-\textbf{r}_k(X_j))}\mu(du)\mu(dv)\Big|\mathcal{D}_k\Big]\Big]
\end{align*}
\begin{align*}
&=\ell\mathbb{E}_{\mathcal{D}_k}\Big[\int|g^*(\textbf{r}_k(u))-\tilde{g}(\textbf{r}_k(u))|^2\times\\
&\quad\ \mathbb{E}_{\{X_j\}_{j=2}^{\ell}}\Big[\int\frac{K_h(\textbf{r}_k(v)-\textbf{r}_k(u))\mu(dv)}{K_h(\textbf{r}_k(v)-\textbf{r}_k(u))+\sum_{j=2}^{\ell}K_h(\textbf{r}_k(v)-\textbf{r}_k(X_j))}\Big|\mathcal{D}_k\Big]\mu(du)\Big]\\
&=\ell\mathbb{E}_{\mathcal{D}_k}\Big[\int|g^*(\textbf{r}_k(u))-\tilde{g}(\textbf{r}_k(u))|^2\times I(u,\ell)\mu(du)\Big].
\end{align*}
Fubini's theorem is employed to obtain the result of the last bound where the inner conditional expectation is denoted by $I(u,\ell)$. We bound $I(u,\ell)$ using the argument of covering $\mathbb{R}^M$ with a countable family of balls $\mathcal{B}\eqdef\{B_M(x_i,\rho/2):i=1,2,....\}$ and the facts that
%$$K_h(\textbf{r}_k(v)-\textbf{r}_k(u))\leq \sum_{i=1}^D K_h(\textbf{r}_k(v)-\textbf{r}_k(u))\mathbbm{1}_{\{B_M(\textbf{r}_k(u)+hx_i,h\rho/2)\}}(\textbf{r}_k(v))$$
\begin{enumerate}
	\item $\textbf{r}_k(v)\in B_M(\textbf{r}_k(u)+hx_i,h\rho/2)\Rightarrow B_M(\textbf{r}_k(u)+hx_i,h\rho/2)\subset B_M(\textbf{r}_k(v),h\rho)$.
	\item $b\mathbbm{1}_{\{B_M(0,\rho)\}}(z) < K(z)\leq 1,\forall z\in\mathbb{R}^M.$
\end{enumerate}

Now, let 
\begin{itemize}
\item $A_{i,h}(u)\eqdef\{v\in\mathbb{R}^d:\|\textbf{r}_k(v)-\textbf{r}_k(u)-hx_i\|<h\rho/2\}$.
\item $B_{i,h}^{\ell}(u)\eqdef\sum_{j=2}^{\ell}\mathbbm{1}_{\{\|\textbf{r}_k(X_j)-\textbf{r}_k(u)-hx_i\|<h\rho/2\}}$. 
\end{itemize}
Thus, one has
%We apply Jensen's inequality of conditional expectation on a convex function of the form $\varphi(x)=\frac{C_0}{C_0+x}$ and obtain,
\begin{align*}
I(u,\ell)&\eqdef\mathbb{E}_{\{X_j\}_{j=2}^{\ell}}\Big[\int\frac{K_h(\textbf{r}_k(v)-\textbf{r}_k(u))\mu(dv)}{K_h(\textbf{r}_k(v)-\textbf{r}_k(u))+\sum_{j=2}^{\ell}K_h(\textbf{r}_k(v)-\textbf{r}_k(X_j))}\Big|\mathcal{D}_k\Big]\\
&\leq\mathbb{E}_{\{X_j\}_{j=2}^{\ell}}\Big[\sum_{i=1}^{+\infty}\int_{v:\|\textbf{r}_k(v)-\textbf{r}_k(u)-hx_i\|<h\rho/2}\\
&\quad\frac{K_h(\textbf{r}_k(v)-\textbf{r}_k(u))\mu(dv)}{K_h(\textbf{r}_k(v)-\textbf{r}_k(u))+\sum_{j=2}^{\ell}K_h(\textbf{r}_k(v)-\textbf{r}_k(X_j))}\Big|\mathcal{D}_k\Big]\\
&\leq\mathbb{E}_{\{X_j\}_{j=2}^{\ell}}\Big[\sum_{i=1}^{+\infty}\int_{A_{i,h}(u)}\\
&\quad\frac{\sup_{z:\|z-hx_i\|<h\rho/2}K_h(z)\mu(dv)}{\sup_{z:\|z-hx_i\|<h\rho/2}K_h(z)+\sum_{j=2}^{\ell}K_h(\textbf{r}_k(v)-\textbf{r}_k(X_j))}\Big|\mathcal{D}_k\Big]\\
&\leq\mathbb{E}_{\{X_j\}_{j=2}^{\ell}}\Big[\sum_{i=1}^{+\infty}\int_{A_{i,h}(u)}\\
&\quad\frac{\sup_{z:\|z-hx_i\|<h\rho/2}K_h(z)\mu(dv)}{\sup_{z:\|z-hx_i\|<h\rho/2}K_h(z)+b\sum_{j=2}^{\ell}\mathbbm{1}_{\{\|\textbf{r}_k(v)-\textbf{r}_k(X_j)\|<h\rho\}}}\Big|\mathcal{D}_k\Big]
\end{align*}
\begin{align*}
&\leq\frac{1}{b}\mathbb{E}_{\{X_j\}_{j=2}^{\ell}}\Big[\sum_{i=1}^{+\infty}\int_{A_{i,h}(u)}\\
&\quad\frac{\sup_{z:\|z-hx_i\|<h\rho/2}K_h(z)\mu(dv)}{\sup_{z:\|z-hx_i\|<h\rho/2}K_h(z)+\sum_{j=2}^{\ell}\mathbbm{1}_{\{\|\textbf{r}_k(X_j)-\textbf{r}_k(u)-hx_i\|<h\rho/2\}}}\Big|\mathcal{D}_k\Big]\\
&\leq\frac{1}{b}\sum_{i=1}^{+\infty}\mathbb{E}_{\{X_j\}_{j=2}^{\ell}}\Big[\frac{\sup_{z:\|z-hx_i\|<h\rho/2}K_h(z)\mu(A_{i,h}(u))}{\sup_{z:\|z-hx_i\|<h\rho/2}K_h(z)+B_{i,h}^{\ell}(u)}\Big|\mathcal{D}_k\Big].
\end{align*}
Note that $B_{i,h}^{\ell}(u)$ is a binomial random variable $B(\ell-1, \mu(A_{i,h}(u)))$ under the law of $\{X_j\}_{j=2}^{\ell}$. Applying part 1 of Lemma 1, one has
\begin{align*}
I(u,\ell)&\leq\frac{1}{b}\sum_{i=1}^{+\infty}\frac{2\sup_{z:\|z-hx_i\|<h\rho/2}K_h(z)\mu(A_{i,h}(u))}{\ell\mu(A_{i,h}(u))}\\
&\leq\frac{2}{b\ell}\sum_{i=1}^{+\infty}\sup_{w:\|w-x_i\|<\rho/2}K(w)\\
&=\frac{2}{b\ell}\sum_{i=1}^{+\infty}\sup_{w\in B_M(x_i,\rho/2)}K(w)\\
&\leq\frac{2}{b\ell}\sum_{i=1}^{+\infty}\sup_{w\in B_M(x_i,\rho/2)}K(w)\\
&\leq\frac{2}{b\ell\lambda_M(B_M(0,\rho/2))}\sum_{i=1}^{+\infty}\int_{B_M(x_i,\rho/2)}\sup_{w\in B_M(x_i,\rho/2)}K(w)dy\\
&\leq\frac{2}{b\ell\lambda_M(B_M(0,\rho/2))}\sum_{i=1}^{+\infty}\int_{B_M(x_i,\rho/2)}\sup_{w\in B_M(y,\rho)}K(w)dy\\
&\leq\frac{2\kappa_M}{b\ell\lambda_M(B_M(0,\rho/2))}\underbrace{\int\sup_{w\in B_M(y,\rho)}K(w)dy}_{=\ \kappa_0\ \text{by \eqref{eq:regular}}}\\
&\leq\frac{2\kappa_M\kappa_0}{b\ell\lambda_M(B_M(0,\rho))}\eqdef\ \frac{C(b,\rho,\kappa_0,M)}{\ell}<+\infty
\end{align*}
where $\lambda_M$ denotes the Lebesque measure on of $\mathbb{R}^M$, $\kappa_M$ denotes the number of balls covering a certain element of $\mathbb{R}^M$, and the constant part is denoted by $C(b,\rho,\kappa_0,M)$ depending on the parameters indicated in the bracket. The last inequality is attained from the fact that the overlapping integrals $\sum_{i=1}^{+\infty}\int_{B_M(x_i,\rho/2)}\sup_{z\in B_M(y,\rho/2)}K(z)dy$ is bounded above by the integral over the entire space $\int\sup_{z\in B_M(y,\rho/2)}K(z)dy$ multiplying by the number of covering balls $k_M$. Therefore,
%\begin{align*}
%I(u,\ell)&\leq\sum_{i\in I_{h,M}}\mu(A_{i,h}(u))\Big[\frac{1}{b}\sup_{z:\|z-x_i\|<\rho/2}K(z)\frac{2}{\ell\mu(A_{i,h}(u))}+\mu(B_{i,h}^{\ell}(u)=0)\Big]\\
%&=\sum_{i\in I_{h,M}}\Big[\frac{2\sup_{z:\|z-x_i\|<\rho/2}K(z)}{b\ell}+\mu(A_{i,h}(u))(\mu([A_{i,h}(u)]^c))^{\ell-1}\Big]\\
%&=\sum_{i\in I_{h,M}}\Big[\frac{2\sup_{z:\|z-x_i\|<\rho/2}K(z)}{b\ell}+\mu(A_{i,h}(u))(1-\mu(A_{i,h}(u)))^{\ell-1}\Big]\\
%&\leq\sum_{i\in I_{h,M}}\Big[\frac{2\sup_{z:\|z-x_i\|<\rho/2}K(z)}{b\ell}+\mu(A_{i,h}(u))e^{-(\ell-1)\mu(A_{i,h}(u))}\Big]
%\end{align*}
%\begin{align*}
%&\leq\frac{2}{b\ell}\sum_{i=1}^{+\infty}\sup_{z:\|z-x_i\|<\rho/2}K(z)+\sum_{i\in I_{h,M}}\mu(A_{i,h}(u))e^{-(\ell-1)\mu(A_{i,h}(u))}\\
%&\leq\frac{2\kappa_0}{b\ell}+\sum_{i\in I_{h,M}}\mu(A_{i,h}(u))e^{-(\ell-1)\mu(A_{i,h}(u))}\\
%&\leq\frac{2\kappa_0}{b\ell}+\sum_{i\in I_h}\mu(A_{i,h}(u))e^{-(\ell-1)\mu(A_{i,h}(u))}\\
%&\leq\frac{2\kappa_0}{b\ell}+\sum_{i\in I_h}\frac{(\ell-1)\mu(A_{i,h}(u))e^{-(\ell-1)\mu(A_{i,h}(u))}}{\ell-1}\\
%&\leq\frac{2\kappa_0}{b\ell}+\frac{|I_h|}{\ell-1}\max_{x\in\mathbb{R}}xe^{-x}\\
%&\leq\frac{2\kappa_0}{b\ell}+\frac{C_0}{h^M}\times\frac{e^{-1}}{\ell-1}.
%\end{align*}
\begin{align*}
A_{n1}&\leq\ell\frac{C(b,\rho,\kappa_0,M)}{\ell}\mathbb{E}_{\mathcal{D}_k}\Big[\int|g^*(\textbf{r}_k(u))-\tilde{g}(\textbf{r}_k(u))|^2\mu(du)\Big]\\
&=C(b,\rho,\kappa_0,M)\mathbb{E}\Big[|\tilde{g}(\textbf{r}_k(X))-g^*(\textbf{r}_k(X))|^2\Big]\\
&<C(b,\rho,\kappa_0,M)\varepsilon.
\end{align*}
%When $\ell\to+\infty$ with $h^M\ell\to+\infty$, the upper bound vanishes and so does $A_{n1}$.
\item Computation of \textbf{$A_{n2}$}: for any $\delta>0$ one has
\begin{align*}
A_{n2}&=\mathbb{E}\Big[\sum_{i=1}^{\ell}W_{n,i}(X)|\tilde{g}(\textbf{r}_k(X_i))-\tilde{g}(\textbf{r}_k(X))|^2\Big]\\
&=\mathbb{E}\Big[\sum_{i=1}^{\ell}W_{n,i}(X)|\tilde{g}(\textbf{r}_k(X_i))-\tilde{g}(\textbf{r}_k(X))|^2\mathbbm{1}_{\{\|\textbf{r}_{k}(X_i)-\textbf{r}_k(X)\|\geq\delta\}}\Big]\\
&\quad+\mathbb{E}\Big[\sum_{i=1}^{\ell}W_{n,i}(X)|\tilde{g}(\textbf{r}_k(X_i))-\tilde{g}(\textbf{r}_k(X))|^2\mathbbm{1}_{\{\|\textbf{r}_{k}(X_i)-\textbf{r}_k(X)\|<\delta\}}\Big]\\
&\leq4\sup_{u\in\mathbb{R}^d}|\tilde{g}(\textbf{r}_k(u))|^2\mathbb{E}\Big[\sum_{i=1}^{\ell}W_{n,i}(X)\mathbbm{1}_{\{\|\textbf{r}_{k}(X_i)-\textbf{r}_k(X)\|\geq\delta\}}\Big]\\
&\quad+ \sup_{u,v\in\mathbb{R}^d:\|\textbf{r}_{k}(u)-\textbf{r}_{k}(v)\|<\delta}|\tilde{g}(\textbf{r}_k(u))-\tilde{g}(\textbf{r}_k(v))|^2
%&\leq4\sup_{u\in\mathbb{R}^d}|\tilde{g}(\textbf{r}_k(u))|^2\mathbb{E}\Big[\sum_{i=1}^{\ell}W_{n,i}(X)\mathbbm{1}_{\{\|\textbf{r}_{k}(X_i)-\textbf{r}_k(X)\|\geq\delta\}}\Big]\\
%&\quad+ \sup_{u,v\in\mathbb{R}^d:\bigwedge\limits_{m=1}^M|r_{k,m}(u)-r_{k,m}(v)|<\delta}|\tilde{g}(\textbf{r}_k(u))-\tilde{g}(\textbf{r}_k(v))|^2.
\end{align*}
Using the uniform continuity of $\tilde{g}$, the second term of the upper bound of $A_{n2}$ tends to $0$ when $\delta$ tends $0$. Thus, we only need to prove that the first term of this upper bound also tends to $0$. We follow a similar procedure as in the previous part:
\begin{align*}
{}&\mathbb{E}\Big[\sum_{i=1}^{\ell}W_{n,i}(X)\mathbbm{1}_{\{\|\textbf{r}_{k}(X_i)-\textbf{r}_k(X)\|\geq\delta\}}\Big]\\
=&\ \mathbb{E}_{\mathcal{D}_k}\Big[\sum_{i=1}^{\ell}\mathbb{E}_{X,\{X_j\}_{j=1}^{\ell}}\Big[W_{n,i}(X)\mathbbm{1}_{\{\|\textbf{r}_{k}(X)-\textbf{r}_k(X_i)\|\geq\delta\}}\Big|\mathcal{D}_k\Big]\Big]\\
=&\ \mathbb{E}_{\mathcal{D}_k}\Big[\sum_{i=1}^{\ell}\mathbb{E}_{\{X_j\}_{j=1}^{\ell}}\Big[\int\frac{K_h(\textbf{r}_k(v)-\textbf{r}_k(X_i))\mathbbm{1}_{\{\|\textbf{r}_{k}(v)-\textbf{r}_k(X_i)\|\geq\delta\}}}{\sum_{j=1}^{\ell}K_h(\textbf{r}_k(v)-\textbf{r}_k(X_j))}\mu(dv)\Big|\mathcal{D}_k\Big]\Big]\\
=&\ \ell\mathbb{E}_{\mathcal{D}_k}\Big[\mathbb{E}_{\{X_j\}_{j=2}^{\ell}}\Big[\int\int\frac{K_h(\textbf{r}_k(v)-\textbf{r}_k(u))\mathbbm{1}_{\{\|\textbf{r}_{k}(v)-\textbf{r}_k(u)\|\geq\delta\}}\mu(du)\mu(dv)}{K_h(\textbf{r}_k(v)-\textbf{r}_k(u))+\sum_{j=2}^{\ell}K_h(\textbf{r}_k(v)-\textbf{r}_k(X_j))}\Big|\mathcal{D}_k\Big]\Big]\\
=&\ \ell\mathbb{E}_{\mathcal{D}_k}\Big[\int J(u,\ell)\mu(du)\Big].
\end{align*}
Fubini's theorem is applied to obtain the last equation where for any $u\in\mathbb{R}^d$,
\begin{align*}
J(u,\ell)&\eqdef\mathbb{E}_{\{X_j\}_{j=2}^{\ell}}\Big[\int\frac{K_h(\textbf{r}_k(v)-\textbf{r}_k(u))\mathbbm{1}_{\{\|\textbf{r}_{k}(v)-\textbf{r}_k(u)\|\geq\delta\}}\mu(dv)}{K_h(\textbf{r}_k(v)-\textbf{r}_k(u))+\sum_{j=2}^{\ell}K_h(\textbf{r}_k(v)-\textbf{r}_k(X_j))}\Big|\mathcal{D}_k\Big]\\
&\leq\mathbb{E}_{\{X_j\}_{j=2}^{\ell}}\Big[\sum_{i=1}^{+\infty}\int_{v:\|\textbf{r}_k(v)-\textbf{r}_k(u)-hx_i\|<h\rho/2}\\
&\quad\ \frac{K_h(\textbf{r}_k(v)-\textbf{r}_k(u))\mathbbm{1}_{\{\|\textbf{r}_{k}(v)-\textbf{r}_k(u)\|\geq\delta\}}}{K_h(\textbf{r}_k(v)-\textbf{r}_k(u))+\sum_{j=2}^{\ell}K_h(\textbf{r}_k(v)-\textbf{r}_k(X_j))}\mu(dv)\Big|\mathcal{D}_k\Big]\\
&\leq\mathbb{E}_{\{X_j\}_{j=2}^{\ell}}\Big[\sum_{i=1}^{+\infty}\int_{A_{i,h}(u)}\\
&\quad\ \frac{\sup_{z:\|z-hx_i\|<h\rho/2}K_h(z)\mathbbm{1}_{\{\|z\|\geq\delta\}}}{\sup_{z:\|z-hx_i\|<h\rho/2}K_h(z)+\sum_{j=2}^{\ell}K_h(\textbf{r}_k(v)-\textbf{r}_k(X_j))}\mu(dv)\Big|\mathcal{D}_k\Big]\\
&\leq\sum_{i=1}^{+\infty}\sup_{z:\|z-hx_i\|<h\rho/2}K_h(z)\mathbbm{1}_{\{\|z\|\geq\delta\}}\times\mathbb{E}_{\{X_j\}_{j=2}^{\ell}}\Big[\int_{A_{i,h}(u)}\\
&\quad\ \frac{\mu(dv)}{\sup_{z:\|z-hx_i\|<h\rho/2}K_h(z)+b\sum_{j=2}^{\ell}\mathbbm{1}_{\{\|\textbf{r}_k(X_j)-\textbf{r}_k(v)\|<h\rho\}}}\Big|\mathcal{D}_k\Big]\\
&\leq\sum_{i=1}^{+\infty}\sup_{z:\|z-hx_i\|<h\rho/2}K_h(z)\mathbbm{1}_{\{\|z\|\geq\delta\}}\times\mathbb{E}_{\{X_j\}_{j=2}^{\ell}}\Big[\int_{A_{i,h}(u)}\\
&\quad\ \mathbb{E}_{\{X_j\}_{j=2}^{\ell}}\Big[\frac{\mu(dv)}{\sup_{z:\|z-hx_i\|<h\rho/2}K_h(z)+b\sum_{j=2}^{\ell}\mathbbm{1}_{\{\|\textbf{r}_k(X_j)-\textbf{r}_k(u)-hx_i\|<h\rho/2\}}}\Big|\mathcal{D}_k\Big]\\
&\leq\sum_{i=1}^{+\infty}\sup_{z:\|z-hx_i\|<h\rho/2}K_h(z)\mathbbm{1}_{\{\|z\|\geq\delta\}}\mu(A_{i,h}(u))\times\\
&\quad\ \frac{1}{b}\mathbb{E}_{\{X_j\}_{j=2}^{\ell}}\Big[\frac{1}{\sup_{z:\|z-hx_i\|<h\rho/2}K_h(z)+B_{i,h}^{\ell}(u)}\Big|\mathcal{D}_k\Big]\\
&\leq\frac{1}{b}\sum_{i=1}^{+\infty}\frac{2\sup_{z:\|z-hx_i\|<h\rho/2}K_h(z)\mu(A_{i,h}(u))\mathbbm{1}_{\{\|z\|\geq\delta\}}}{\ell \mu(A_{i,h}(u))}\\
&\leq\frac{2}{b\ell}\sum_{i=1}^{+\infty}\sup_{w:\|w-x_i\|<\rho/2}K(w)\mathbbm{1}_{\{\|w\|\geq\delta/h\}}.
\end{align*}
Thus, one has
\begin{align*}
\mathbb{E}\Big[\sum_{i=1}^{\ell}W_{n,i}(X)\mathbbm{1}_{\{\|\textbf{r}_{k}(X_i)-\textbf{r}_k(X)\|\geq\delta\}}\Big]
&\leq \ell\frac{2}{b\ell}\sum_{i=1}^{+\infty}\sup_{w\in B_M(x_i,\rho/2)}K(w)\mathbbm{1}_{\{\|w\|\geq\delta/h\}}
\end{align*}
When both $h\to0$ and $\delta\to0$ satisfying $\delta/h\to+\infty$, the upper bound series converges to zero. Indeed, it is a non-negative convergent series thanks to the proof of $I(u,l)$ in the previous part. Moreover, the general term of the series, $s_k=\sup_{w\in B_M(x_k,\rho/2)}K(w)\mathbbm{1}_{\{\|w\|\geq\delta/h\}}$, satisfying $\lim_{\delta/h\to+\infty} s_k=0$ for all $k\geq 1$. Therefore, this series converges to zero when $h\to0,\delta\to0$ such that $\delta/h\to+\infty$.
\end{itemize}
In conclusion, when $\ell\to+\infty$ and $\varepsilon,h,\delta\to0$ such that $\delta/h\to+\infty$, all the three terms of the upper bound of $A_n$ tend to $0$, so does $A_n$.

\QED
\end{proofA}
% ================= Proposition A.2 ===================

\begin{propA}
Under the assumptions of \textbf{Proposition~2},
\begin{align*}
\lim_{\ell\rightarrow+\infty}\mathbb{E}\Big[\Big|\sum_{i=1}^{\ell}W_{n,i}(X)[Y_i-g_n(\textbf{r}_k(X_i))]\Big|^2\Big]=0.
\end{align*}
\end{propA}

\begin{proofA} Using the independence between $(X_i,Y_i)$ and $(X_j,Y_j)$ for all $i\neq j$, one has 
\begin{align*}
A.2&=\mathbb{E}\Big[\Big|\sum_{i=1}^{\ell}W_{n,i}(X)[Y_i-g_n(\textbf{r}_k(X_i))]\Big|^2\Big]\\
&=\sum_{1\leq i,j\leq\ell}\mathbb{E}\Big[W_{n,i}(X)W_{n,j}(X)[Y_i-g_n(\textbf{r}_k(X_i))][Y_j-g_n(\textbf{r}_k(X_j))]\Big]\\
&=\ \mathbb{E}\Big[\sum_{i=1}^{\ell}W_{n,i}^2(X)|Y_i-g_n(\textbf{r}_k(X_i))|^2\Big]=\mathbb{E}\Big[\sum_{i=1}^{\ell}W_{n,i}^2(X)\sigma^2(\textbf{r}_k(X_i))\Big]
\end{align*}
where 
$$\sigma^2(\textbf{r}_k(x))\eqdef\mathbb{E}[(Y_i-g_n(\textbf{r}_k(X_i)))^2|\textbf{r}_k(x)].$$
Thus, based on the assumption of $X$ and $Y$ we have $\sigma^2\in L_1(\mu)$. Therefore, $\sigma^2$ can be approximated in $L_1$ sense i.e., for any $\varepsilon>0, \exists\tilde{\sigma}^2$ a continuous function with compact support such that
$$\mathbb{E}[|\sigma^2(\textbf{r}_k(X))-\tilde{\sigma}^2(\textbf{r}_k(X))|]<\varepsilon.$$
Thus, one has
\begin{align*}
A.2&\leq\mathbb{E}\Big[\sum_{i=1}^{\ell}W_{n,i}^2(X)\tilde{\sigma}^2(\textbf{r}_k(X_i))\Big]+\mathbb{E}\Big[\sum_{i=1}^{\ell}W_{n,i}^2(X)|\sigma^2(\textbf{r}_k(X_i))-\tilde{\sigma}^2(\textbf{r}_k(X_i))|\Big]\\
&\leq\sup_{u\in\mathbb{R}^d}|\tilde{\sigma}^2(\textbf{r}_k(u))|\mathbb{E}\Big[\sum_{i=1}^{\ell}W_{n,i}^2(X)\Big]+\mathbb{E}\Big[\sum_{i=1}^{\ell}W_{n,i}^2(X)|\sigma^2(\textbf{r}_k(X_i))-\tilde{\sigma}^2(\textbf{r}_k(X_i))|\Big].
\end{align*}
Using similar argument as in the case of $A_{n1}$ and the fact that $W_{n,i}(x)\leq 1,\forall i=1,2,...,\ell$, thus for any $\varepsilon>0$, one has
\begin{align*}
\mathbb{E}\Big[\sum_{i=1}^{\ell}W_{n,i}^2(X)|\sigma^2(\textbf{r}_k(X_i))-\tilde{\sigma}^2(\textbf{r}_k(X_i))|\Big]&\leq\mathbb{E}\Big[\sum_{i=1}^{\ell}W_{n,i}(X)|\sigma^2(\textbf{r}_k(X_i))-\tilde{\sigma}^2(\textbf{r}_k(X_i))|\Big]\\
&<C(b,\rho,\kappa_0,M)\varepsilon.
\end{align*}
Therefore, it remains to prove that $\mathbb{E}[\sum_{i=1}^{\ell}W_{n,i}^2(X)]\to0$ as $\ell\to+\infty$. As $b\mathbbm{1}_{\{B_M(0,\rho)\}}(z) < K(z)\leq 1,\forall z\in\mathbb{R}^M$ with the convention of $0/0=0$, for a fixed $\delta>0$, one has
%$\sum_{j=1}^{\ell}K_h(\textbf{r}_k(X)-\textbf{r}_k(X_j))>0$
\begin{align}
\sum_{i=1}^{\ell}W_{n,i}^2(X)&=\sum_{i=1}^{\ell}\Big(\frac{K_h(\textbf{r}_k(X)-\textbf{r}_k(X_i))}{\sum_{j=1}^{\ell}K_h(\textbf{r}_k(X)-\textbf{r}_k(X_j))}\Big)^2\nonumber\\
&\leq\frac{\sum_{i=1}^{\ell}K_h(\textbf{r}_k(X)-\textbf{r}_k(X_i))}{\Big(\sum_{j=1}^{\ell}K_h(\textbf{r}_k(X)-\textbf{r}_k(X_j))\Big)^2}\nonumber\\
&\leq\min\Big\{\delta,\frac{\mathbbm{1}_{\{\sum_{j=1}^{\ell}K_h(\textbf{r}_k(X)-\textbf{r}_k(X_j))>0\}}}{\sum_{j=1}^{\ell}K_h(\textbf{r}_k(X)-\textbf{r}_k(X_j))}\Big\}\nonumber\\
&\leq\min\Big\{\delta,\frac{\mathbbm{1}_{\{\sum_{j=1}^{\ell}\mathbbm{1}_{\{\|\textbf{r}_k(X)-\textbf{r}_k(X_j)\|<h\rho\}}>0\}}}{b\sum_{j=1}^{\ell}\mathbbm{1}_{\{\|\textbf{r}_k(X)-\textbf{r}_k(X_j)\|<h\rho\}}}\nonumber\\
&\leq\delta+\frac{\mathbbm{1}_{\{\sum_{j=1}^{\ell}\mathbbm{1}_{\{\|\textbf{r}_k(X)-\textbf{r}_k(X_j)\|<h\rho\}}>0\}}}{b\sum_{j=1}^{\ell}\mathbbm{1}_{\{\|\textbf{r}_k(X)-\textbf{r}_k(X_j)\|<h\rho\}}}.\label{eq:boundW2}
\end{align}
Therefore, it is enough to show that
$$\mathbb{E}\Big[\frac{\mathbbm{1}_{\{\sum_{j=1}^{\ell}\mathbbm{1}_{\{\|\textbf{r}_k(X)-\textbf{r}_k(X_j)\|<h\rho\}}>0\}}}{\sum_{j=1}^{\ell}\mathbbm{1}_{\{\|\textbf{r}_k(X)-\textbf{r}_k(X_j)\|<h\rho\}}}\Big]\xrightarrow{\ell\to+\infty}0.$$
One has
\begin{align*}
{}&\ \ \ \ \mathbb{E}\Big[\frac{\mathbbm{1}_{\{\sum_{j=1}^{\ell}\mathbbm{1}_{\{\|\textbf{r}_k(X)-\textbf{r}_k(X_j)\|<h\rho\}}>0\}}}{\sum_{j=1}^{\ell}\mathbbm{1}_{\{\|\textbf{r}_k(X)-\textbf{r}_k(X_j)\|<h\rho\}}}\Big]\\
&\leq\mathbb{E}\Big[\frac{\mathbbm{1}_{\{\sum_{j=1}^{\ell}\mathbbm{1}_{\{\|\textbf{r}_k(X)-\textbf{r}_k(X_j)\|<h\rho\}}>0\}}}{\sum_{j=1}^{\ell}\mathbbm{1}_{\{\|\textbf{r}_k(X)-\textbf{r}_k(X_j)\|<h\rho\}}}\mathbbm{1}_{\{\textbf{r}_k(X)\in B\}}\Big]+\mu(\{v\in\mathbb{R}^d:\textbf{r}_k(v)\in B^c\})\\
&=\mathbb{E}\Big[\mathbbm{1}_{\{\textbf{r}_k(X)\in B\}}\mathbb{E}\Big[\frac{\mathbbm{1}_{\{\sum_{j=1}^{\ell}\mathbbm{1}_{\{\|\textbf{r}_k(X)-\textbf{r}_k(X_j)\|<h\rho\}}>0\}}}{\sum_{j=1}^{\ell}\mathbbm{1}_{\{\|\textbf{r}_k(X)-\textbf{r}_k(X_j)\|<h\rho\}}}\Big|X\Big]\Big]+\mu(\{v\in\mathbb{R}^d:\textbf{r}_k(v)\in B^c\})\\
&\leq2\mathbb{E}\Big[\frac{\mathbbm{1}_{\{\textbf{r}_k(X)\in B\}}}{(\ell+1)\mu(\{v\in\mathbb{R}^d:\|\textbf{r}_k(v)-\textbf{r}_k(X)\|<h\rho\})}\Big]+\mu(\{v\in\mathbb{R}^d:\textbf{r}_k(v)\in B^c\})
\end{align*}
where $B$ is a $M$-dimensional ball centered at the origin chosen so that the second term $\mu(\{v\in\mathbb{R}^d:\textbf{r}_k(v)\in B^c\})$ is small. The last inequality is attained by applying part 2 of lemma 1. Moreover, as $\textbf{r}_k=(\textbf{r}_{k,m})_{m=1}^M$ is bounded then there exists a finite number of balls in $\mathcal{B}=\{B_M(x_j,h\rho/2):j=1,2,...\}$ such that $B$ is contained in the union of these balls i.e., $\exists I_{h,M}$ finite, such that $B\subset\cup_{j\in I_{h,M}}B_M(x_j,h\rho/2)$. 
\begin{align}
{}&\quad\ \mathbb{E}\Big[\frac{\mathbbm{1}_{\{\textbf{r}_k(X)\in B\}}}{(\ell+1)\mu(\{v\in\mathbb{R}^d:\|\textbf{r}_k(v)-\textbf{r}_k(X)\|<h\rho\})}\Big]\nonumber\\
&\leq\sum_{j\in I_{h,M}}\int_{u:\|\textbf{r}_k(u)-x_j\|<h\rho/2}\frac{\mu(du)}{(\ell+1)\mu(\{v\in\mathbb{R}^d:\|\textbf{r}_k(v)-\textbf{r}_k(u)\|<h\rho\})}\nonumber\\
&\quad+\mu(\{v\in\mathbb{R}^d:\textbf{r}_k(v)\in B^c\})\nonumber\\
&\leq\sum_{j\in I_{h,M}}\int_{u:\|\textbf{r}_k(u)-x_j\|<h\rho/2}\frac{\mu(du)}{(\ell+1)\mu(\{v\in\mathbb{R}^d:\|\textbf{r}_k(v)-x_j\|<h\rho/2\})}\nonumber\\
&\quad+\mu(\{v\in\mathbb{R}^d:\textbf{r}_k(v)\in B^c\})\nonumber\\
&=\sum_{j\in I_{h,M}}\frac{\mu(\{u\in\mathbb{R}^d:\|\textbf{r}_k(u)-x_j\|<h\rho/2\})}{(\ell+1)\mu(\{v\in\mathbb{R}^d:\|\textbf{r}_k(v)-x_j\|<h\rho/2\})}+\mu(\{v\in\mathbb{R}^d:\textbf{r}_k(v)\in B^c\})\nonumber\\
&=\frac{|I_{h,M}|}{\ell+1}+\mu(\{v\in\mathbb{R}^d:\textbf{r}_k(v)\in B^c\})\nonumber\\
&\leq\frac{C_0}{h^M(\ell+1)}+\mu(\{v\in\mathbb{R}^d:\textbf{r}_k(v)\in B^c\})\label{eq:bound1}\\
&\xrightarrow[h^M\ell\to+\infty]{\ell\to+\infty,h\to0}\mu(\{v\in\mathbb{R}^d:\textbf{r}_k(v)\in B^c\}). \nonumber
\end{align}
It is easy to check the following fact,
\begin{equation}
\label{eq:boundJhM}
|I_{h,M}|\leq\frac{C_0}{h^M}\ \text{for some }C_0>0.
\end{equation}
To prove this, we consider again the cover $\mathcal{B}=\{B_M(x_j,h\rho/2):j=1,2,...\}$ of $\mathbb{R}^M$. For any $\rho>0$ fixed and $h>0$, note that
the covering number $|I_{h,M}|$ is proportional to the ratio between the volume of $B$ and the volume of the ball $B_M(0,h\rho/2)$ i.e.,
\begin{align*}
|I_{h,M}|&\propto \frac{\text{Vol}(B)}{\text{Vol}(B_M(0,h\rho/2))}\\
&\propto \frac{\text{Vol}(B)}{(h\rho/2)^M}\\
&\leq  \frac{C_0}{h^M}
\end{align*}
for some positive constant $C_0$ proportional to the volume of $B$. Finally, we can conclude the proof of the proposition as we can choose $B$ such that $\mu(\{v\in\mathbb{R}^d:\textbf{r}_k(v)\in B^c\})=0$ thanks to the boundedness of the basic machines.
\begin{remark}
The assumption on the boundedness of the constructed machines is crucial. This assumption allows us to choose a ball B which can be covered using a finite number $|I_{h,M}|$ of balls $B_M(x_j,h\rho/2)$, therefore makes it possible to prove the result of this proposition for this class of regular kernels. Note that for the class of compactly supported kernels, it is easy to obtain such a result directly from the begging of the evaluation of each integral (see, for example, Chapter 5 of \cite{bookDistributionFree}).
\end{remark}

\QED
\end{proofA}
\begin{propA}
Under the assumptions of \textbf{Proposition~2},
\begin{align*}
\lim_{\ell\rightarrow+\infty}\mathbb{E}\Big[\Big|g^*(\textbf{r}_k(X))\Big(\sum_{i=1}^{\ell}W_{n,i}(X)-1\Big)\Big|^2\Big]=0.
\end{align*}
\end{propA}
%===============================================
\begin{proofA}
Note that $|\sum_{i=1}^{\ell}W_{n,i}(X)-1|\leq 1$ thus one has
$$\Big|g^*(\textbf{r}_k(X))\Big(\sum_{i=1}^{\ell}W_{n,i}(X)-1\Big)\Big|^2\leq|g^*(\textbf{r}_k(X))|^2.$$
Consequently, by Lebesque's dominated convergence theorem, to prove this proposition, it is enough to show that $\sum_{i=1}^{\ell}W_{n,i}(X)\to1$ almost surely.
Note that $1-\sum_{i=1}^{\ell}W_{n,i}(X)=\mathbbm{1}_{\{\sum_{i=1}^{\ell}K_h(\textbf{r}_k(X)-\textbf{r}_k(X_i))=0\}}$ therefore,
\begin{align*}
\mathbb{P}\Big[\sum_{i=1}^{\ell}W_{n,i}(X)\neq 1\Big]&=\mathbb{P}\Big[\sum_{i=1}^{\ell}K_h(\textbf{r}_k(X)-\textbf{r}_k(X_i))=0\Big]\\
&\leq\mathbb{P}\Big(\sum_{j=1}^{\ell}\mathbbm{1}_{\{\|\textbf{r}_k(X)-\textbf{r}_k(X_j)\|<h\rho\}}=0\Big)\\
&= \int\mathbb{P}\Big(\sum_{j=1}^{\ell}\mathbbm{1}_{\{\|\textbf{r}_k(x)-\textbf{r}_k(X_j)\|<h\rho\}}=0\Big)\mu(dx)\\
&= \int\mathbb{P}\Big(\cap_{j=1}^{\ell}\{\|\textbf{r}_k(x)-\textbf{r}_k(X_j)\|\geq h\rho\}\Big)\mu(dx)\\
&= \int\Big[1-\mathbb{P}\Big(\{\|\textbf{r}_k(x)-\textbf{r}_k(X_1)\|< h\rho\}\Big)\Big]^{\ell}\mu(dx)\\
&= \int\Big[1-\mu\Big(\{v\in\mathbb{R}^d:\|\textbf{r}_k(x)-\textbf{r}_k(v)\|< h\rho\}\Big)\Big]^{\ell}\mu(dx)
\end{align*}
\begin{align*}
&\leq \int e^{-\ell\mu(A_h(x))}\mu(dx)\\
&=\int e^{-\ell\mu(A_h(x))}\mathbbm{1}_{\{\textbf{r}_k(x)\in B\}}\mu(dx)+\mu(\{v\in\mathbb{R}^d:\textbf{r}_k(v)\in B^c\})\\
&\leq\frac{\max_{u}\{ue^{-u}\}}{\ell}\int \frac{\mathbbm{1}_{\{\textbf{r}_k(x)\in B\}}}{\mu(A_h(x))}\mu(dx)+\mu(\{v\in\mathbb{R}^d:\textbf{r}_k(v)\in B^c\})
\end{align*}
where
\begin{equation}
\label{eq:Axh}
A_h(x)\eqdef\{v\in\mathbb{R}^d:\|\textbf{r}_k(x)-\textbf{r}_k(v)\|< h\rho\}.
\end{equation}
Therefore,
\begin{align*}
\mathbb{P}\Big[\sum_{i=1}^{\ell}W_{n,i}(X)\neq 1\Big]&\leq\frac{e^{-1}}{\ell}\mathbb{E}\Big[\frac{\mathbbm{1}_{\{\textbf{r}_k(X)\in B\}}}{\mu(\{v\in\mathbb{R}^d:\|\textbf{r}_k(v)-\textbf{r}_k(X)\|<h\rho\})}\Big]\\
&\quad+\mu(\{v\in\mathbb{R}^d:\textbf{r}_k(v)\in B^c\}).
\end{align*}
Following the same procedure as in the proof of $A.2$ we obtain the desire result.
\end{proofA}
\QED
\end{proofprop}

%=========== Theorem 1 ===========
\begin{proofthm}
Choose a new observation $x\in\mathbb{R}^d$, given the training data $\mathcal{D}_k$ and the predictions $\{\textbf{r}_k(X_p)\}_{p=1}^{\ell}$ on $\mathcal{D}_{\ell}$, taking expectation with respect to the response variables $\{Y_p^{(\ell)}\}_{p=1}^{\ell}$, it is easy to check that
\begin{align*}
&{}\ \ \ \ \mathbb{E}[|g_n(\textbf{r}_k(x))-g^*(\textbf{r}_k(x))|^2|\{\textbf{r}_k(X_p)\}_{p=1}^{\ell},\mathcal{D}_k]\\
&=\mathbb{E}\Big[\Big|g_n(\textbf{r}_k(x))-\mathbb{E}[g_n(\textbf{r}_k(x))|\{\textbf{r}_k(X_p)\}_{p=1}^{\ell},\mathcal{D}_k]\\
&\quad+\mathbb{E}[g_n(\textbf{r}_k(x))|\{\textbf{r}_k(X_p)\}_{p=1}^{\ell},\mathcal{D}_k]-g^*(\textbf{r}_k(x))\Big|^2\Big|\{\textbf{r}_k(X_p)\}_{p=1}^{\ell},\mathcal{D}_k\Big]\\
&=\mathbb{E}[|g_n(\textbf{r}_k(x))-\mathbb{E}[g_n(\textbf{r}_k(x))|\{\textbf{r}_k(X_p)\}_{p=1}^{\ell},\mathcal{D}_k]|^2|\{\textbf{r}_k(X_p)\}_{p=1}^{\ell},\mathcal{D}_k]\\
&\quad+|g^*(\textbf{r}_k(x))-\mathbb{E}[g_n(\textbf{r}_k(x))|\{\textbf{r}_k(X_p)\}_{p=1}^{\ell},\mathcal{D}_k]|^2\\
&\eqdef E_1+E_2.
\end{align*}
On one hand by using the independence between $Y_i$ and $(Y_j,X_j)$ for all $i\neq j$, we develop the square and obtain for any $\delta>0$:

\begin{align*}
E_1&\eqdef\mathbb{E}\Big[\Big|g_n(\textbf{r}_k(x))-\mathbb{E}[g_n(\textbf{r}_k(x))|\{\textbf{r}_k(X_p)\}_{p=1}^{\ell},\mathcal{D}_k]\Big|^2\Big|\{\textbf{r}_k(X_p)\}_{p=1}^{\ell},\mathcal{D}_k\Big]\\
&=\mathbb{E}\Big[\Big|\sum_{i=1}^{\ell}W_{n,i}(x)(Y_i-\mathbb{E}[Y_i|\textbf{r}_k(X_i)])\Big|^2\Big|\{\textbf{r}_k(X_p)\}_{p=1}^{\ell},\mathcal{D}_k\Big]\\
&=\mathbb{E}\Big[\sum_{i=1}^{\ell}W_{n,i}^2(x)(Y_i-\mathbb{E}[Y_i|\textbf{r}_k(X_i)])^2\Big|\{\textbf{r}_k(X_p)\}_{p=1}^{\ell},\mathcal{D}_k\Big]\\
&=\sum_{i=1}^{\ell}W_{n,i}^2(x)\mathbb{E}_{Y_i}[(Y_i-\mathbb{E}[Y_i|\textbf{r}_k(X_i)])^2|\textbf{r}_k(X_i)]\\
&=\mathbb{V}[Y_1|\textbf{r}_k(X_1)]\sum_{i=1}^{\ell}W_{n,i}^2(x)\\
&\overset{(\ref{eq:boundW2})}{\leq} \frac{4R^2}{b}\Big(\delta+\frac{\mathbbm{1}_{\{\sum_{j=1}^{\ell}\mathbbm{1}_{\{\|\textbf{r}_k(x)-\textbf{r}_k(X_j)\|<h\rho\}}>0\}}}{\sum_{j=1}^{\ell}\mathbbm{1}_{\{\|\textbf{r}_k(x)-\textbf{r}_k(X_j)\|<h\rho\}}}\Big)
\end{align*}
where the notation $\mathbb{V}(Z)$ stands for the variance of a random variable $Z$. Therefore, using the result of inequality~(\ref{eq:bound1}), one has
\begin{equation}
\label{eq:boundE1}
\mathbb{E}(E_1)\leq \frac{4R^2}{b}\Big(\delta+\frac{C_0}{h^M(\ell+1)}\Big)
\end{equation}
for some $C_0>0$. On the other hand, set 
\begin{itemize}
\item[--] $C_{h}^{\ell}(x)\eqdef\sum_{j=1}^{\ell}\mathbbm{1}_{\{\|\textbf{r}_k(X_j)-\textbf{r}_k(x)\|<h\rho\}}.$
\item[--] $D_{h}^{\ell}(x)\eqdef\sum_{j=1}^{\ell}K_h({r}_k(X_j)-\textbf{r}_k(x)).$
\end{itemize}
The second term $E_2$ is hard to control as it depends on the behavior of $g^*(\textbf{r}_k(.))$. That is why a weak smoothness assumption of the theorem is required to connect this behavior to the behavior of the input machines. Using this assumption, one has
\begin{align*}
E_2&\eqdef\Big|g^*(\textbf{r}_k(x))-\mathbb{E}[g_n(\textbf{r}_k(x))|\{\textbf{r}_k(X_p)\}_{p=1}^{\ell},\mathcal{D}_k]\Big|^2\\
&=\Big(\sum_{i=1}^{\ell}W_{n,i}(X)(g^*(\textbf{r}_k(x))-\mathbb{E}[Y_i|\textbf{r}_k(X_i)])\Big)^2\mathbbm{1}_{\{D_h^{\ell}(x)>0\}}\\
&\quad+(g^*(\textbf{r}_k(x)))^2\mathbbm{1}_{\{D_h^{\ell}(x)=0\}}
\end{align*}
\begin{align*}
&\overset{(\text{Jensen})}{\leq\ \ \ \ \ }\sum_{i=1}^{\ell}W_{n,i}(x)(g^*(\textbf{r}_k(x))-\mathbb{E}[Y_i|\textbf{r}_k(X_i)])^2\mathbbm{1}_{\{D_h^{\ell}(x)>0\}}\nonumber\\
&\quad+(g^*(\textbf{r}_k(x)))^2\mathbbm{1}_{\{D_h^{\ell}(x)=0\}}\\
&\leq\sum_{i=1}^{\ell}\frac{K_h(\textbf{r}_k(x)-\textbf{r}_k(X_i))(g^*(\textbf{r}_k(x))-g^*(\textbf{r}_k(X_i)))^2}{\sum_{j=1}^{\ell}K_h(\textbf{r}_k(x)-\textbf{r}_k(X_j))}\mathbbm{1}_{\{D_h^{\ell}(x)>0\}}\nonumber\\
&\quad+(g^*(\textbf{r}_k(x)))^2\mathbbm{1}_{\{D_h^{\ell}(x)=0\}}\\
&\leq L^2\sum_{i=1}^{\ell}\frac{K_h(\textbf{r}_k(x)-\textbf{r}_k(X_i))\|\textbf{r}_k(x)-\textbf{r}_k(X_i)\|^2}{\sum_{j=1}^{\ell}K_h(\textbf{r}_k(x)-\textbf{r}_k(X_j))}\mathbbm{1}_{\{D_h^{\ell}(x)>0\}}\nonumber\\
&\quad+(g^*(\textbf{r}_k(x)))^2\mathbbm{1}_{\{D_h^{\ell}(x)=0\}}\\
&\leq L^2\Big[\sum_{i=1}^{\ell}\frac{K_h(\textbf{r}_k(x)-\textbf{r}_k(X_i))\|\textbf{r}_k(x)-\textbf{r}_k(X_i)\|^2\mathbbm{1}_{\{\|\textbf{r}_k(x)-\textbf{r}_k(X_i)\|<R_Kh^{\alpha}\}}}{\sum_{j=1}^{\ell}K_h(\textbf{r}_k(x)-\textbf{r}_k(X_j))}\\
&\quad+\sum_{i=1}^{\ell}\frac{K_h(\textbf{r}_k(x)-\textbf{r}_k(X_i))\|\textbf{r}_k(x)-\textbf{r}_k(X_i)\|^2\mathbbm{1}_{\{\|\textbf{r}_k(x)-\textbf{r}_k(X_i)\|\geq R_Kh^{\alpha}\}}}{\sum_{j=1}^{\ell}K_h(\textbf{r}_k(x)-\textbf{r}_k(X_j))}\Big]\times\\
&\quad\quad \mathbbm{1}_{\{D_h^{\ell}(x)>0\}}+(g^*(\textbf{r}_k(x)))^2\mathbbm{1}_{\{C_h^{\ell}(x)=0\}}\\
&\eqdef E_2^1+E_2^2+E_2^3.
\end{align*}
for any $\alpha\in(0,1)$ chosen arbitrarily at this point. Now, we bound the expectation of the three terms of the last inequality. 

\begin{itemize}
\item Firstly, $E_2^1$ can be easily bounded from above by
\begin{align*}
E_2^1&\eqdef L^2\sum_{i=1}^{\ell}\frac{K_h(\textbf{r}_k(x)-\textbf{r}_k(X_i))\|\textbf{r}_k(x)-\textbf{r}_k(X_i)\|^2}{\sum_{j=1}^{\ell}K_h(\textbf{r}_k(x)-\textbf{r}_k(X_j))}\mathbbm{1}_{\{D_h^{\ell}(x)>0\}}\times\\
&\quad\ \mathbbm{1}_{\{\|\textbf{r}_k(x)-\textbf{r}_k(X_i)\|<R_Kh^{\alpha}\}}\\
&\leq L^2h^{2\alpha}R_K^2\sum_{i=1}^{\ell}\frac{K_h(\textbf{r}_k(x)-\textbf{r}_k(X_i))}{\sum_{j=1}^{\ell}K_h(\textbf{r}_k(x)-\textbf{r}_k(X_j))}\mathbbm{1}_{\{D_h^{\ell}(x)>0\}}\\
&= L^2h^{2\alpha}R_K^2.
\end{align*}
Therefore, its expectation is simply bounded by the same upper bound i.e.,
\begin{align}
\label{eq:boundE21}
\mathbb{E}(E_2^1)\leq L^2h^{2\alpha}R_K^2
\end{align}
\item Secondly, we bound the second term $E_2^2$ using the tail assumption of the kernel $K$ given equation~\eqref{eq:assumption}, thus for any $h>0$:
\begin{align*}
E_2^2&\eqdef h^2L^2\sum_{i=1}^{\ell}\frac{K_h(\textbf{r}_k(x)-\textbf{r}_k(X_i))\|(\textbf{r}_k(x)-\textbf{r}_k(X_i))/h\|^2}{\sum_{j=1}^{\ell}K_h(\textbf{r}_k(x)-\textbf{r}_k(X_j))}\mathbbm{1}_{\{D_h^{\ell}(x)>0\}}\times\\
&\quad\ \mathbbm{1}_{\{\|\textbf{r}_k(x)-\textbf{r}_k(X_i)\|\geq R_Kh^{\alpha}\}}\\
&\leq h^2L^2\sum_{i=1}^{\ell}\frac{C_K\mathbbm{1}_{\{\|(\textbf{r}_k(x)-\textbf{r}_k(X_i))/h\|\geq R_K/h^{1-\alpha}\}}\mathbbm{1}_{\{D_h^{\ell}(x)>0\}}}{(1+\|(\textbf{r}_k(x)-\textbf{r}_k(X_i))/h\|^{M})\sum_{j=1}^{\ell}K_h(\textbf{r}_k(x)-\textbf{r}_k(X_j))}\\
&\leq h^{M+2}L^2\sum_{i=1}^{\ell}\frac{C_K\mathbbm{1}_{\{\|\textbf{r}_k(x)-\textbf{r}_k(X_i)\|\geq R_Kh^{\alpha}\}}\mathbbm{1}_{\{D_h^{\ell}(x)>0\}}}{(h^{M}+(R_Kh^{\alpha})^{M})\sum_{j=1}^{\ell}K_h(\textbf{r}_k(x)-\textbf{r}_k(X_j))}\\
&\leq h^{M+2}L^2C_K\sum_{i=1}^{\ell}\frac{\mathbbm{1}_{\{\|\textbf{r}_k(x)-\textbf{r}_k(X_i)\|\geq R_Kh^{\alpha}\}}}{(h^{M}+R_K^{M}h^{\alpha M})\sum_{j=1}^{\ell}K_h(\textbf{r}_k(x)-\textbf{r}_k(X_j))}\mathbbm{1}_{\{D_h^{\ell}(x)>0\}}\nonumber\\
&\leq \frac{h^{M+2-\alpha M}L^2C_K}{h^{M(1-\alpha)}+R_K^{M}}\times\frac{\sum_{i=1}^{\ell}\mathbbm{1}_{\{\|\textbf{r}_k(x)-\textbf{r}_k(X_i)\|\geq R_Kh^{\alpha}\}}}{\sum_{j=1}^{\ell}K_h(\textbf{r}_k(x)-\textbf{r}_k(X_j))}\mathbbm{1}_{\{D_h^{\ell}(x)>0\}}\\
&\leq\frac{h^{(1-\alpha)M+2}L^2C_K}{bR_K^{M}}\times\frac{\sum_{i=1}^{\ell}\mathbbm{1}_{\{\|\textbf{r}_k(x)-\textbf{r}_k(X_i)\|\geq R_Kh^{\alpha}\}}}{\sum_{j=1}^{\ell}\mathbbm{1}_{\{\|\textbf{r}_k(x)-\textbf{r}_k(X_j)\|<h\rho\}}}\mathbbm{1}_{\{C_h^{\ell}(x)>0\}}.
\end{align*}
Therefore,
\begin{align*}
E_2^2&\leq\frac{h^{(1-\alpha)M+2}L^2C_K\ell}{bR_K^{M}}\times\frac{\mathbbm{1}_{\{\sum_{j=1}^{\ell}\mathbbm{1}_{\{\|\textbf{r}_k(x)-\textbf{r}_k(X_j)\|<h\rho\}}>0\}}}{\sum_{j=1}^{\ell}\mathbbm{1}_{\{\|\textbf{r}_k(x)-\textbf{r}_k(X_j)\|<h\rho\}}}.
\end{align*}
Again, applying the result of inequality~(\ref{eq:bound1}), one has
\begin{align}
\label{eq:boundE22}
\mathbb{E}(E_2^2)\leq\frac{h^{(1-\alpha)M+2}L^2C_K\ell}{bR_K^{M}}\times\frac{C_0}{h^M(\ell+1)}\leq \frac{C_1\ell}{(\ell+1)}h^{2-\alpha M}
\end{align}
for some $C_1>0$ and $\alpha < 2/M$.
\item Lastly with $A_h(x)$ defined in (\ref{eq:Axh}), we bound the expectation of $E_2^3$ by,
\begin{align*}
\mathbb{E}(E_2^3)&\leq \mathbb{E}\Big[(g^*(\textbf{r}_k(x)))^2\mathbbm{1}_{\{C_h^{\ell}(x)=0\}}\Big]\\
&\leq \sup_{u\in\mathbb{R}^d}(g^*(\textbf{r}_k(u)))^2\mathbb{E}\Big[\mathbbm{1}_{\{C_h^{\ell}(x)=0\}}\Big]\\
&=\sup_{u\in\mathbb{R}^d}(g^*(\textbf{r}_k(u)))^2(1-\mu(A_h(x)))^{\ell}\\
&\leq \sup_{u\in\mathbb{R}^d}(g^*(\textbf{r}_k(u)))^2e^{-\ell\mu(A_h(x))}
\end{align*}
\begin{align}
&\leq \sup_{u\in\mathbb{R}^d}(g^*(\textbf{r}_k(u)))^2\frac{\ell\mu(A_h(x))e^{-\ell\mu(A_h(x))}}{\ell\mu(A_h(x))}\nonumber\\
&\leq \sup_{u\in\mathbb{R}^d}(g^*(\textbf{r}_k(u)))^2\frac{\max_{u\in\mathbb{R}^d}ue^{-u}}{\ell\mu(A_h(x))}\nonumber\\
&\leq \sup_{u\in\mathbb{R}^d}(g^*(\textbf{r}_k(u)))^2\frac{e^{-1}}{\ell\mu(A_h(x))}\nonumber\\
&\leq \frac{C_2}{\ell\mu(A_h(x)))}\label{eq:boundE23}
\end{align}
for some $C_2>0$.
\end{itemize}
From \eqref{eq:boundE1}, \eqref{eq:boundE21}, \eqref{eq:boundE22} and \eqref{eq:boundE23}, one has
\begin{align*}
\mathbb{E}[|g_n(\textbf{r}_k(X))-g^*(\textbf{r}_k(X))|^2]&\leq\int_{\mathbb{R}^d}\mathbb{E}[|g_n(\textbf{r}_k(x))-g^*(\textbf{r}_k(x))|^2]\mu(dx) \\
&\leq \int_{\mathbb{R}^d}\mathbb{E}(E_1+E_2^1+E_2^2+E_2^3)\mu(dx)\\
&\leq \int_{\mathbb{R}^d}\Big[\frac{4R^2}{b}\Big(\delta+\frac{C_0}{h^M(\ell+1)}\Big) + L^2h^{2\alpha}R_K^2 \\
&\quad+ \frac{C_1\ell}{(\ell+1)}h^{2-\alpha M} + \frac{C_2}{\ell\mu(A_h(x)))}\Big]\mu(dx).
\end{align*}
Therefore following the same procedure of proving inequality~(\ref{eq:bound1}), one has
\begin{align*}
&\quad\mathbb{E}[|g_n(\textbf{r}_k(X))-g^*(\textbf{r}_k(X))|^2]\\
&\leq \frac{4R^2}{b}\Big(\delta+\frac{C_0}{h^M(\ell+1)}\Big) + L^2h^{2\alpha}R_K^2+\frac{C_1\ell}{(\ell+1)}h^{2-\alpha M} + \int_{\mathbb{R}^d}\frac{C_2\mu(dx)}{\ell\mu(A_h(x)))}\\
&\leq \frac{4R^2}{b}\Big(\delta+\frac{C_0}{h^M(\ell+1)}\Big)+ L^2h^{2\alpha}R_K^2+\frac{C_1\ell}{(\ell+1)}h^{2-\alpha M} \\
&\quad+\sum_{j\in J_{h,M}}\int_{\|\textbf{r}_k(x)-x_j\|<h\rho}\frac{C_2\mu(dx)}{\ell\mu(\{v\in\mathbb{R}^d:\|\textbf{r}_k(v)-\textbf{r}_k(x)\|<h\rho\})}\\
&\leq \frac{4R^2}{b}\Big(\delta+\frac{C_0}{h^M(\ell+1)}\Big)+ L^2h^{2\alpha}R_K^2+\frac{C_1\ell}{(\ell+1)}h^{2-\alpha M} \\
&\quad+\sum_{j\in J_{h,M}}\int_{\|\textbf{r}_k(x)-x_j\|<h\rho}\frac{C_2\mu(dx)}{\ell\mu(\{v\in\mathbb{R}^d:\|\textbf{r}_k(v)-x_j\|<h\rho\})}
\end{align*}
\begin{align*}
&\leq \frac{4R^2}{b}\Big(\delta+\frac{C_0}{h^M(\ell+1)}\Big)+ L^2h^{2\alpha}R_K^2+\frac{C_1\ell}{(\ell+1)}h^{2-\alpha M} \\
&\quad+\frac{C_2}{\ell}\sum_{j\in J_{h,M}}\frac{\mu(\{v\in\mathbb{R}^d:\|\textbf{r}_k(v)-x_j\|<h\rho\})}{\mu(\{v\in\mathbb{R}^d:\|\textbf{r}_k(v)-x_j\|<h\rho\})}\\
&\leq\frac{4R^2}{b}\Big(\delta+\frac{C_0}{h^M(\ell+1)}\Big)+ L^2h^{2\alpha}R_K^2+\frac{C_1\ell}{(\ell+1)}h^{2-\alpha M}+\frac{C_2|J_{h,M}|}{\ell}\\
&\leq\frac{4R^2}{b}\Big(\delta+\frac{C_0}{h^M(\ell+1)}\Big)+ L^2R_K^2h^{2\alpha}+\frac{C_1\ell}{(\ell+1)}h^{2-\alpha M}+\frac{C_2'}{h^M\ell}
\end{align*}
where $|J_{h,M}|$ denotes the number of balls covering the ball $B$ (introduced in the proof of $A.2$) by the cover $\{B_M(x_j,h\rho):j=1,2,...\}$. Similarly, one has $|J_{h,M}|\leq \frac{C_0}{h^M}$ for some constant $C_0>0$ proportional to the volume of $B$. Since $\delta>0$ can be arbitrarily small, and with the choice of $\alpha=2/(M+2)$, we can deduce that
\begin{equation}
\label{eq:final}
\mathbb{E}[|g_n(\textbf{r}_k(X))-g^*(\textbf{r}_k(X))|^2]\leq \frac{\tilde{C}_1}{h^M\ell}+\tilde{C}_2h^{4/(M+2)}.
\end{equation}
From this bound, for $h\propto \ell^{-(M+2)/(M^2+2M+4)}$ we obtain the desire result with the upper bound of order $O(\ell^{-\frac{4}{M^2+2M+4}})$ i.e.,
$$\mathbb{E}[|g_n(\textbf{r}_k(X))-g^*(\textbf{r}_k(X))|^2]\leq C\ell^{-\frac{4}{M^2+2M+4}}$$
for some constant $C>0$ independent of $\ell$.
\end{proofthm}
\QED
\begin{proofremark}
To prove the result in this case, which means, under the following assumption:
$$\exists R_K,C_K>0\text{ and }\alpha\in(0,1):K(z)\leq C_Ke^{-\|z\|^{\alpha}},\forall z\in\mathbb{R}^M,\|z\|\geq R_K,$$
we only need to check the new bound of $E_2^2$ defined in the previous case. One has
\begin{align*}
E_2^2&\eqdef L^2\sum_{i=1}^{\ell}\frac{K_h(\textbf{r}_k(x)-\textbf{r}_k(X_i))\|\textbf{r}_k(x)-\textbf{r}_k(X_i)\|^2\mathbbm{1}_{\{D_h^{\ell}(x)>0\}}}{\sum_{j=1}^{\ell}K_h(\textbf{r}_k(x)-\textbf{r}_k(X_j))}\times\\
&\quad\ \mathbbm{1}_{\{\|\textbf{r}_k(x)-\textbf{r}_k(X_i)\|\geq h^{\alpha}R_K\}}\\
&\leq L^2\sum_{i=1}^{\ell}\frac{h^2K_h(\textbf{r}_k(x)-\textbf{r}_k(X_i))\|(\textbf{r}_k(x)-\textbf{r}_k(X_i))/h\|^2\mathbbm{1}_{\{D_h^{\ell}(x)>0\}}}{\sum_{j=1}^{\ell}K_h(\textbf{r}_k(x)-\textbf{r}_k(X_j))}\\
&\quad\ \mathbbm{1}_{\{(\|\textbf{r}_k(x)-\textbf{r}_k(X_i))/h\|\geq R_K/h^{1-\alpha}\}}
\end{align*}
\begin{align*}
&\leq \frac{h^2L^2}{b}\sum_{i=1}^{\ell}\frac{C_Ke^{-\|(\textbf{r}_k(x)-\textbf{r}_k(X_i))/h\|^{\alpha}}\|(\textbf{r}_k(x)-\textbf{r}_k(X_i))/h\|^2}{\sum_{j=1}^{\ell} \mathbbm{1}_{\{\|\textbf{r}_k(x)-\textbf{r}_k(X_j)\|<h\rho\}}}\times\\
&\quad\ \mathbbm{1}_{\{\|(\textbf{r}_k(x)-\textbf{r}_k(X_i))/h\|\geq R_K/h^{1-\alpha}\}}\mathbbm{1}_{\{C_h^{\ell}(x)>0\}}
\end{align*}
As for any $\alpha\in(0,1)$, $t\mapsto \lambda(t)=t^2e^{-t^{\alpha}}$ is strictly decreasing for all $t\geq (2/\alpha)^{1/\alpha}$. Thus for $h$ small enough such that $R_K/h^{1-\alpha}\geq(2/\alpha)^{1/\alpha}$, one has

\begin{align*}
E_2^2&\leq \frac{h^2L^2C_K}{b}\sum_{i=1}^{\ell}\frac{(R_K/h^{1-\alpha})^2e^{-(R_K/h^{1-\alpha})^{\alpha}}\mathbbm{1}_{\{\|(\textbf{r}_k(x)-\textbf{r}_k(X_i))/h\|\geq R_K/h^{1-\alpha}\}}}{\sum_{j=1}^{\ell} \mathbbm{1}_{\{\|\textbf{r}_k(x)-\textbf{r}_k(X_j)\|<h\rho\}}}\mathbbm{1}_{\{C_h^{\ell}(x)>0\}}\\
&\leq \frac{h^{2\alpha}L^2C_KR_K^2e^{-R_K^{\alpha} h^{-\alpha(1-\alpha)}}}{b}\sum_{i=1}^{\ell}\frac{\mathbbm{1}_{\{\sum_{j=1}^{\ell} \mathbbm{1}_{\{\|\textbf{r}_k(x)-\textbf{r}_k(X_j)\|<h\rho\}}>0\}}}{\sum_{j=1}^{\ell} \mathbbm{1}_{\{\|\textbf{r}_k(x)-\textbf{r}_k(X_j)\|<h\rho\}}}\\
&\leq \frac{\ell h^{2\alpha}L^2C_KR_K^2e^{-R_K^{\alpha} h^{-\alpha(1-\alpha)}}}{b}\times\frac{\mathbbm{1}_{\{\sum_{j=1}^{\ell}\mathbbm{1}_{\{\|\textbf{r}_k(x)-\textbf{r}_k(X_j)\|<h\rho\}}>0\}}}{\sum_{j=1}^{\ell}\mathbbm{1}_{\{\|\textbf{r}_k(x)-\textbf{r}_k(X_j)\|<h\rho\}}}.
\end{align*}
Applying the result of inequality~(\ref{eq:bound1}), one has
\begin{align}
\mathbb{E}(E_2^2)&\leq\frac{\ell h^{2\alpha}L^2C_KR_K^2e^{-R_K^{\alpha} h^{-\alpha(1-\alpha)}}}{b}\times\frac{C_0}{h^M(\ell+1)}\nonumber\\
&\leq C_1h^{2\alpha-M}e^{-R_K^{\alpha}h^{-\alpha(1-\alpha)}}\label{eq:boundE22.1}
\end{align}
for some $C_1>0$ and $\alpha\in(0,1)$.

\noindent Therefore from \eqref{eq:boundE1}, \eqref{eq:boundE21}, \eqref{eq:boundE23} and \eqref{eq:boundE22.1}, one has
\begin{align*}
\mathbb{E}[|g_n(\textbf{r}_k(X))-g^*(\textbf{r}_k(X))|^2]&\leq\int_{\mathbb{R}^d}\mathbb{E}[|g_n(\textbf{r}_k(x))-g^*(\textbf{r}_k(x))|^2]\mu(dx) \\
&\leq \int_{\mathbb{R}^d}\mathbb{E}(E_1+E_2^1+E_2^2+E_2^3)\mu(dx)\\
&\leq \int_{\mathbb{R}^d}\Big[\frac{4R^2}{b}\Big(\delta+\frac{C_0}{h^M(\ell+1)}\Big) + L^2h^{2\alpha}R_K^2 \\
&\quad+ C_1h^{2\alpha-M}e^{-R_K^{\alpha}h^{-\alpha(1-\alpha)}} + \frac{C_2}{\ell\mu(A_h(x)))}\Big]\mu(dx).
\end{align*}
Following the same procedure as in the previous proof of theorem 1, one has
\begin{align*}
&\quad\mathbb{E}[|g_n(\textbf{r}_k(X))-g^*(\textbf{r}_k(X))|^2]\\
&\leq\frac{4R^2}{b}\Big(\delta+\frac{C_0}{h^M(\ell+1)}\Big)+ L^2h^{2\alpha}R_K^2+C_1h^{2\alpha-M}e^{-R_K^{\alpha}h^{-\alpha(1-\alpha)}}+\frac{C_2'}{h^M\ell}.
\end{align*}
Since $\delta>0$ is chosen arbitrarily and the third term of the last inequality decreases exponentially fast as $h\to0$ for any $\alpha\in(0,1)$ fixed, hence it is negligible comparing to other terms. Finally, with the choice of $h\propto \ell^{-1/(M+2\alpha)}$, we obtain the desire result:
\begin{equation*}
\label{eq:final2}
\mathbb{E}[|g_n(\textbf{r}_k(X))-g^*(\textbf{r}_k(X))|^2]\leq \frac{\tilde{C}_1}{h^M\ell}+\tilde{C}_2h^{2\alpha}\leq C\ell^{-2\alpha/(M+2\alpha)}
\end{equation*}
for some $C>0$ independent of $\ell$.
\end{proofremark}
\QED
%\bibliographystyle{elsarticle-harv}
%\nocite{*}
%\bibliography{bibs-KCOBRA}

\end{document}